\RequirePackage{rotating}
\documentclass[fleqn,usenatbib]{mnras}

\usepackage{newtxtext,newtxmath}

\usepackage[T1]{fontenc}
\usepackage{rotating}
\usepackage{comment}

\DeclareRobustCommand{\VAN}[3]{#2}
\let\VANthebibliography\thebibliography
\def\thebibliography{\DeclareRobustCommand{\VAN}[3]{##3}\VANthebibliography}

\usepackage{graphicx}	
\usepackage{amsmath}	
\usepackage{amsthm}
\usepackage{subfigure}
\usepackage{amssymb}	
\usepackage{xcolor}
\usepackage{orcidlink}
\usepackage{CJKutf8}
\usepackage{adjustbox}
\usepackage{lscape}
\usepackage{multirow}

\newcommand{\Mdust}{$M_{\rm dust}$}	
\newcommand{\Mgas}{$M_{\rm gas}$}
\newcommand{\MHI}{$M_{\rm HI}$}
\newcommand{\Mmol}{$M_{\rm mol}$}
\newcommand{\Matom}{$M_{\rm atom}$}
\newcommand{\Mstar}{$M_{\star}$}
\newcommand{\Zgas}{$Z_{\rm gas}$}
\newcommand{\Zsol}{$Z_{\rm \odot}$}
\newcommand{\Msol}{$M_{\rm \odot}$}
\newcommand{\logOH}{12+log(O/H)}
\newcommand{\kms}{km~s$^{-1}$}
\newcommand{\alphacomw}{$\alpha_{\rm CO, MW}$}
\newcommand{\alphacoz}{$\alpha_{\rm CO, Z}$}
\newcommand{\MHt}{$M_{\rm H_2}$}

\title[Resolved dust evolution in nearby galaxies]{The spatially resolved relation between dust, gas, and metal abundance with the TYPHOON survey}

\author[H.-J. Park et al.]{
    Hye-Jin Park$^{\orcidlink{0000-0002-9809-6631}}$,$^{1,2}$\thanks{E-mail: hyejin.park@anu.edu.au}
    Andrew J. Battisti$^{\orcidlink{0000-0003-4569-2285}}$,$^{1,2}$
    Emily Wisnioski$^{\orcidlink{0000-0003-1657-7878}}$, $^{1,2}$
    Luca Cortese$^{\orcidlink{0000-0002-7422-9823}}$,$^{2,3}$
    Mark Seibert$^{\orcidlink{0000-0002-1143-5515}}$, $^{4}$
    \newauthor
    Kathryn Grasha$^{\orcidlink{0000-0002-3247-5321}}$,$^{1,2,5}$\thanks{ARC DECRA Fellow},
    Barry F. Madore$^{\orcidlink{0000-0002-1576-1676}}$,$^{4,6}$
    Brent Groves$^{\orcidlink{0000-0002-9768-0246}}$, $^{2,3}$
    Jeff A. Rich$^{\orcidlink{0000-0002-5807-5078}}$, $^{4}$
    Rachael L. Beaton$^{\orcidlink{0000-0002-1691-8217}}$, $^{4, 7}$\thanks{Hubble Fellow}
    \newauthor
    Qian-Hui Chen (陈千惠)$^{\orcidlink{0000-0002-4382-1090}}$,$^{1,2}$
    Marcie Mun$^{\orcidlink{0000-0002-3706-9955}}$,$^{1,2}$
    Naomi M. McClure-Griffiths$^{\orcidlink{0000-0003-2730-957X}}$,$^{1}$
    W.J.G. de Blok$^{\orcidlink{0000-0001-8957-4518}}$,$^{8,9,10}$
    \newauthor
    Lisa J. Kewley$^{\orcidlink{0000-0001-8152-3943}}$ $^{11,1,2}$
    \\ 
    $^{1}$Research School of Astronomy and Astrophysics, Australian National University, Cotter Road, Weston Creek, ACT 2611, Australia\\
    $^{2}$ARC Centre of Excellence for All Sky Astrophysics in 3 Dimensions (ASTRO 3D), Australia\\
    $^{3}$International Centre for Radio Astronomy Research, The University of Western Australia, 35 Stirling Highway, Crawley, Perth, WA 6009, Australia\\
    $^{4}$The Observatories, Carnegie Institution for Science, 813 Santa Barbara Street, Pasadena, CA 91101, USA\\
    $^{5}$Visiting Fellow, Harvard-Smithsonian Center for Astrophysics, 60 Garden Street, Cambridge, MA 02138, USA\\
    $^{6}$Department of Astronomy and Astrophysics, University of Chicago, Chicago, IL, USA\\
    $^{7}$Department of Astrophysical Sciences, 4 Ivy Lane, Princeton University, Princeton, NJ 08544, USA\\
    $^{8}$ASTRON, the Netherlands Institute for Radio Astronomy, Oude Hoogeveensedijk 4, 7991 PD Dwingeloo, the Netherlands\\
    $^{9}$Department of Astronomy, University of Cape Town, Private Bag X3, Rondebosch 7701, South Africa\\
    $^{10}$Kapteyn Astronomical Institute, University of Groningen, PO Box 800, 9700 AV Groningen, The Netherlands\\
    $^{11}$Institute for Theory and Computation, Harvard-Smithsonian Center for Astrophysics, Cambridge, MA 02138, USA\\
}

\date{Accepted XXX. Received YYY; in original form ZZZ}

\pubyear{2024}

\begin{document}
\label{firstpage}
\pagerange{\pageref{firstpage}--\pageref{lastpage}}
\begin{CJK}{UTF8}{gbsn}
	\label{firstpage}
	\pagerange{\pageref{firstpage}--\pageref{lastpage}}
	\maketitle

\begin{abstract}
We present the spatially resolved relationship between the dust-to-gas mass ratio (DGR) and gas-phase metallicity (\Zgas\ or \logOH) (i.e., DGR\---\Zgas\ relation) of 11 nearby galaxies with a large metallicity range (1.5~dex of \logOH) at (sub-)kpc scales. We used the large field-of-view ($\gtrsim$ 3\arcmin) optical pseudo-Integral Field Spectroscopy data taken by the TYPHOON/PrISM survey, covering the optical size of galaxies, combining them with multi-wavelength data (far-UV to far-IR, CO, and H{\sc i} 21~cm radio). A large scatter of DGR in the intermediate metallicity galaxies (8.0 $<$ \logOH $<$ 8.3) is found, which is in line with dust evolution models, where grain growth begins to dominate the mechanism of dust mass accumulation. In the lowest metallicity galaxy of our sample, Sextans A (\logOH $<$ 7.6), the star-forming regions have significantly higher DGR values (by 0.5\-–2 dex) than the global estimates from literature at the same metallicity but aligns with the DGR values from metal depletion method from Damped Lyman Alpha systems and high hydrogen gas density regions of Sextans A. Using dust evolution models with a Bayesian MCMC approach suggests: 1) a high SN dust yield and 2) a negligible amount of photofragmentation by UV radiation, although we note that our sample in the low-metallicity regime is limited to Sextans A. On the other hand, it is also possible that while metallicity influences DGR, gas density also plays a role, indicating an early onset of dust grain growth in the dust mass build-up process despite its low metallicity. 

\end{abstract}




\section{Introduction}
\label{sec:introduction}

Interstellar dust plays an important role in the interstellar medium (ISM) of galaxies and understanding this role is fundamental to improving our theories of galaxy evolution. Despite its small mass fraction out of the total baryon mass ($\leq$\,1\% in a typical late-type galaxy; \citealt{galliano2018interstellar}), interstellar dust has a disproportionate effect on a galaxy's ISM, including: being catalysts of chemical reactions on the surface of dust grains, such as the formation of molecular hydrogen (H$_2$) (\citealt{hollenbach1971surface}; \citealt{hollenbach1997dense}) and being major coolants of the interstellar medium (ISM) by shielding them from the interstellar radiation field, aiding the collapse of the giant molecular clouds (GMC) into forming new stars. Dust accretes and removes chemical elements from the gas phase, known as `metal\footnote{Heavy elements compared to hydrogen and helium.} depletion.' The \textit{selective} depletion of elements onto dust, known as `fractionation,' impacts the gas phase, changing the observed gas abundances and removing potential coolants. In addition, dust significantly affects the spectral energy distribution (SED) of galaxies and introduces uncertainties in derived physical properties, such as the star formation rate (SFR) and stellar mass (\Mstar) (\citealt{conroy2013modeling}). Specifically, the starlight emitted from young massive (O/B\--type) stars at short wavelengths is absorbed by interstellar dust and re-emitted at longer wavelengths. Generally speaking, roughly 30~\--~50~\% of the light from young stellar populations is lost at wavelengths from far-UV to near-IR and re-emitted at mid-IR to sub-mm wavelengths (\citealt{galliano2018interstellar}).

The dust-to-gas mass ratio (DGR),
\begin{equation}
    DGR \equiv M_{\rm dust} / M_{\rm gas} = M_{\rm dust} / (M_{\rm atom} + M_{\rm mol}),
    \label{eq:dgr}
\end{equation}
can represent how much a region is enriched, in the form of dust mass (\Mdust) at a given amount of gas mass (the sum of atomic gas mass and molecular gas mass, \Mgas = \Matom + \Mmol). This mass ratio is a reflection of various sources contributing to dust growth and destruction. Assuming the equilibrium between timescales of dust formation and dust destruction within galaxies, a tight linear relationship between the DGR and the gas-phase metallicity (\Zgas\ or \logOH\ in this study) of galaxies is expected (e.g., \citealt{inoue2011origin}; \citealt{asano2013dust}), and it has been observed in many metal-rich galaxies (\Zgas\ $>$~0.45~\Zsol or \logOH $>$ 8.3; e.g., \citealt{issa1990dust}, \citealt{devis2017using}, \citealt{devis2019}). 

The linear DGR\---\Zgas\ relation is a useful tool itself in estimating gas masses of massive (and metal-rich) high-redshift ($z$) galaxies via \Mdust\ because measuring \Mdust\ via dust continuum in Rayleigh-Jeans (RJ) tail (e.g., \citealt{casey2014dusty}) and converting it to gas mass is more observationally efficient compared to measuring gas mass directly from the CO or H{\sc i} observations. For example, \citet{scoville2014evolution} and \citet{scoville2016ism} have developed a calibration for \Mgas estimation using the dust continuum for 107 galaxies with high stellar mass (\Mstar\ $\sim$~$10^{11}$\Msol) in the COSMOS field at the $z=0.2-2.5$. They constructed the calibrations by assuming constant DGR values at solar metallicity. More recently, \citet{tacconi2018phibss} provided an updated calibration to estimate gas mass using the empirical trend of a linearly increasing DGR with gas-phase metallicity for \logOH~$>$~8.0. However, these calibrations are often widely used across a range of high-$z$ galaxies (e.g., \citealt{liu2019automated}) despite it being unclear how reliable these extrapolations are down to the low mass/metallicity regime.

\citet{remyruyer2014} studied the relationship between galaxy-integrated DGR and \Zgas\ ({\it global DGR\---\Zgas\ relation} hereafter) using 126 nearby galaxies from two surveys, Key Insights on Nearby Galaxies: A Far-Infrared Survey with Herschel (KINGFISH; \citealt{kennicutt2011kingfish}) and Dwarf Galaxy Survey (DGS; \citealt{madden2013overview}) of 7.14 $<$ \logOH $<$ 9.10 (with metallicity diagnostic using $R_{\rm 23}$ ratio\footnote{$R_{\rm 23}$ = ([O~II]$\lambda$ 3726,3729 + [O~III]$\lambda$ 4959,5007) / H$\beta$)} of \citealt{pilyugin2005oxygen}; PT05). They found that the observed global DGR\---\Zgas\ relation is better described with a broken power law rather than a single power law, with a steeper slope and a larger scatter at low metallicity galaxies (\logOH $\lesssim$ 8.0). They suggested that the break at \logOH $\mathrm \sim 8.0$ implies that the balance of timescales of dust formation/growth and destruction is not preserved in low metallicity systems. However, more recently \citet{devis2019} with 466 DustPedia late-type galaxies (\citealt{davies2017dustpedia}) found that a single power law better describes the DGR trend with \Zgas\ over the broken model, redirecting the exact shape of DGR\---\Zgas\ relation on the debate. 

Theoretical approaches, on the other hand, show the steeper slope of the DGR\---\Zgas\ relation at lower metallicity (e.g. \citealt{zhukovska2014dust}, \citealt{feldmann2015equilibrium}, \citealt{devis2017using}, and \citealt{galliano2021nearby}), implying the non-equilibrium between dust formation and destruction timescales and different mechanisms on the dust mass build-up. In particular, \citet{asano2013dust} suggested `critical metallicity' where the dominant mechanism of dust mass increase is changed and this value can vary with the star formation timescale of a system. They further divide the metallicity ($Z$) into three regimes: 
\begin{enumerate}
    \item[-] Low-metallicity regime (starburst regime; $Z$ $<$ 0.05 \Zsol): dust formation or evolution is slow as dust grains are mainly condensed in stellar ejecta (stellar wind or SNe).
    \item[-] Intermediate-metallicity regime (critical metallicity regime; 0.05\---0.3 \Zsol): dust mass build-up starts to be efficient as grain growth in the ISM comes into play as a dominance.
    \item[-] High-metallicity regime (ISM growth regime; $Z$ $>$ 0.3 \Zsol): grain growth dominates, showing a linear relation of DGR $\propto$ $Z$, consistent with many previous observational studies in the high-metallicity regime.
\end{enumerate}

The spatially resolved study on the DGR\---\Zgas\ relation is important as \Mdust\ measurement at galaxy-integrated scale, as probed in the most earlier works, can be underestimated in SED modelling due to the resolution effect (also known as Matryoshka effect; \citealt{galliano2011non} and \citealt{galliano2018interstellar}). This is because hot dust components can dilute the luminosity of colder components due to cold dust's weak luminosity (e.g., modified black body) at a coarse resolution. In terms of accurate \Mdust\ measurement of a system, this effect is crucial as cold dust constitutes a substantial fraction of the total \Mdust. This can lead to a systematically underestimated total \Mdust\ by up to 50~\% when it is derived in a galaxy-integrated way (e.g., \citealt{galliano2011non}, \citealt{galliano2018interstellar}, Appendix D of \citealt{roman-duval2014dust}, \citealt{aniano2012modeling}, and \citealt{aniano2020modeling}).

Owing to the development of optical IFS, it has become possible to derive the DGR\---\Zgas\ relation in a spatially resolved manner ({\it resolved DGR\---\Zgas\ relation} hereafter). However, most resolved DGR\---\Zgas\ relations have been derived with a fixed metallicity gradient for individual galaxies and are mostly focused on high-metallicity regimes with smaller Field-of-View (FoV) only covering the central region of galaxies. For example, \citet{sandstrom2013co} explored the spatially resolved DGR analysis (at $\sim$~kpc scale) for 26 nearby galaxies with a metallicity range of 8.2 $<$ \logOH\ (PT05) $<$ 8.75. The metallicity values of each region rely on the metallicity gradient from a small part of galaxies rather than the entire galaxy (\citealt{moustakas2010optical}). More recently, several studies such as \citet{vilchez2018metals}, \citet{relano2018spatially}, and \citet{chiang2018spatially} explored the resolved DGR\---\Zgas\ relation in nearby spiral galaxies in a similar way (i.e. fixed metallicity gradient). However, assuming a metallicity gradient of galaxies can overlook the locally varying metal abundances in galaxies, which can be related to the age of the stellar cluster, molecular gas surface density, dust column density, and electron density (e.g., \citealt{kreckel2019mapping}; \citealt{groves2023phangs}). This indicates that fully spatially resolved studies on this relation are required to mitigate these effects.

In this study, we explore the DGR\---\Zgas\ relationship by examining \textit{both} of DGR and \Zgas\ in a fully spatially resolved way using a sample of 11 nearby galaxies from the TYPHOON survey that spans a broad range in metallicity (7.1 $<$ \logOH\ (Scal) $<$ 8.6).

The outline of this paper consists of as follows. We describe our galaxy sample, the datasets, and data processing in Section~\ref{sec:sample_data}. In Section~\ref{sec:physicalparam}, we present the method for the measurements of ISM parameters for resolved regions. We show the derived global and resolved DGR\---\Zgas\ relation and several ISM scaling relations in Section~\ref{sec:result}. In Section~\ref{sec:discussion}, we discuss the tension in the DGR\---\Zgas\ relationship and comparison to dust/chemical evolution models. We then summarize our conclusions in Section~\ref{sec:summary}. Throughout the paper, we assumes the flat $\Lambda$-CDM cosmological model with $H_0$ = 70 \kms~Mpc$^{-1}$, $\Omega_M$ = 0.3, and $\Omega_\Lambda$ = 0.7.

\section{Sample and data}
\label{sec:sample_data}

\subsection{TYPHOON Survey}
\label{sec:typhoon} 

The TYPHOON/PrISM survey\footnote{\url{https://typhoon.datacentral.org.au}} (Carnegie Observatories, TYPHOON Programme PI: Barry F. Madore) provides pseudo-IFS data with a large Field-of-View (FoV) size for 44 nearby galaxies (D $\lesssim$ 35~Mpc) with a wide range in morphological types such as dwarf irregulars, spirals, ring, and AGN/Seyfert. This survey used the Wide-field CCD imaging spectrograph of the du Pont 2.5m telescope at Las Campanas Observatory in Chile. The data are obtained with a high physical resolution ($\sim$~100~pc/spaxel at a distance of 12.5~Mpc derived from a spatial resolution of $\sim$~1.65\arcsec) and a spectral resolution of R~$\sim$~850 at 7000~\AA. The Progressive Integral Step Method (PrISM) is a `stepped-slit' technique to create 3D data cubes. The observations have a large FoV of approximately 18\arcmin $\times$\,(1.65\arcsec $\times \rm N$), where N represents the number of stepped slits. This enables coverage of nearly the entire optical disk of very nearby galaxies (e.g., \citealt{poetrodjojo2019effects}, \citealt{grasha2022metallicity}, \citealt{chen2023metallicity}, and Siebert et al., in prep). This dataset offers a unique opportunity to gain a comprehensive understanding of the distributions of physical and chemical properties across the extended source and also allows us to explore many galaxy evolution studies such as the metallicity gradient and bar effect (e.g., \citealt{grasha2022metallicity} and \citealt{chen2023metallicity}).

The target galaxies of the TYPHOON survey were selected from the sample of The 11~Mpc Halpha and Ultraviolet Galaxy Survey (11HUGS) and the Local Volume Legacy (LVL) surveys. The TYPHOON parent sample galaxies are located on the declination of $< + 10~\deg$ and have low to moderate inclination and a surface brightness brighter than 21.1 AB mag arcsec$^{-2}$ in the B band. For this study, we select our sample galaxies based on the availability of ancillary data ({\it Herschel} far-IR, CO (1-0) or CO (2-1), and H{\sc i} 21~cm). The physical properties of the 11 TYPHOON galaxies that satisfy these ancillary data requirements are summarised in Table~\ref{tab:sample}. Further details on the individual characteristics of the galaxies are presented in Appendix~\ref{app:eachsample}. 

We provide a brief description of the TYPHOON data in this paper (please refer to the forthcoming project description paper, Seibert et al. in prep., for detailed information on the data and its reduction process). The final data cubes, obtained after data reduction, have the following characteristics: (1) wavelength coverage ranging from 3650\---8150~\AA, (2) spectral resolution of $\Delta\lambda$$\sim$~3.5~\AA\ (FWHM of 8.24~\AA), corresponding to R~$\sim$~850 at 7000~\AA\ and (3) spatial resolution of 1.65\arcsec, which corresponds to a physical scale of 40~pc at a representative distance of 5~Mpc. The emission line fluxes of the spaxels are fitted using \texttt{lzifu} (\citealt{ho2016lzifu}; Battisti et al. in prep). After modelling and subtracting the stellar continuum from each observed spectra, \texttt{lzifu} fits a single Gaussian to the emission line and measures its flux. The following emission lines are extracted: [O~II]$\lambda$3726,\,3729, H$\delta$, H$\gamma$, [O~III]$\lambda$4363, H$\beta$, [O~III]$\lambda$5007, H$\alpha$, [N~II]$\lambda$6583, [S~II]$\lambda$6716,~6731.

\begin{table*} 
    \centering
    \begin{tabular}{c|c|c|c|c|c|c|c|c|c|c|c|c}
    \hline
    \hline
    Name     & RA & DEC & Dist. & $z$ & Phys. Resol. & $r_{25}$ & P.A. & Incl. & log \Mstar & log SFR & A(V)$_{\rm MW}$$^*$ \\
             & (J2000) & (J2000) & (Mpc) & & (kpc) & (arcsec) & (deg) & (deg) & ($M_{\odot}$) & ($M_{\odot}$ yr$^{-1}$) & (mag) \\
    \hline
    NGC~625  & 01h35m04.63s & -41d26m10.3s & 3.5$^{a}$ & 0.001321 & 0.424 & 172.7 & 106$^{1}$ & 64$^{1}$ & 8.48$^{6}$ & -1.40$^{6}$ & 0.0437 \\
    NGC~1512 & 04h03m54.28s & -43d20m55.9s & 18.8$^{b}$ & 0.002995 & 2.279 & 267.4 & 261.9$^{2}$ & 42.5$^{2}$ & 10.57$^{5}$ & -0.12$^{5}$ & 0.0282\\
    NGC~1566 & 04h20m00.42s	& -54d56m16.1s & 17.7$^{c}$ & 0.005017 & 2.145 & 249.6 & 214.7$^{2}$ & 29.5$^{2}$ & 10.67$^{5}$ & 0.65$^{5}$ & 0.0242\\
    NGC~1705 & 04h54m13.50s	& -53d21m39.8s & 5.5$^{a}$ & 0.002112 & 0.667 & 57.2 & 50$^{1}$ & 45$^{1}$ & 8.07$^{5}$ & -1.29$^{5}$ & 0.0213 \\
    NGC~3521 & 11h05m48.58s & -00d02m09.1s & 13.2$^{d}$ & 0.002672 & 1.600 & 329.0 & 343$^{2}$ & 68.8$^{2}$ & 10.83$^{5}$ & 0.42$^{5}$ & 0.153\\
    NGC~4536 & 12h34m27.05s & +02d11m17.3s & 16.3$^{e}$ & 0.006031 & 1.976 & 227.6 & 305.6$^{2}$ & 66$^{2}$ & 10.19$^{5}$ & 0.47$^{5}$ & 0.0487\\
    NGC~5236 & 13h37m00.95s & -29d51m55.5s & 4.9$^{e}$ & 0.001711 & 0.594 & 386.5 & 225$^{2}$ & 24$^{2}$ & 10.41$^{5}$ & 0.63$^{5}$ & 0.1770 \\
    NGC~5253 & 13h39m55.96s	& -31d38m24.4s & 3.5$^{a}$ & 0.001358 & 0.424 & 150.4 & 21$^{1}$ & 42$^{1}$ & 8.57$^{5}$ & -0.32$^{5}$ & 0.1481 \\
    NGC~6822 & 19h44m57.74s & -14d48m12.4s & 0.5$^{a}$ & -0.00019 & 0.200 & 464.7 & 10$^{3}$$^{\dagger}$ & 67$^{3}$$^{\dagger}$  & 7.90$^{6}$ & -2.15$^{6}$ & 0.6174\\
    NGC~7793 & 23h57m49.83s & -32d35m27.7s & 3.6$^{e}$ & 0.000767 & 0.436 & 280.0 & 290$^{2}$ & 50$^{2}$ & 9.25$^{5}$ & -0.60$^{5}$ & 0.0518\\
    Sextans~A & 10h11m00.80s & -04d41m34.0s & 1.4$^{a}$ & 0.001081 & 0.200 & 176.7 & 41$^{4}$ & 33.5$^{4}$ & 8.04$^{5}$ & -2.18$^{5}$ & 0.1198 \\
    \hline
    \hline
    \end{tabular}
    \caption{Characteristics of sample galaxies. The coordinates and redshifts are from NED. The chosen physical resolution of each galaxy in this study after data processing (convolution and resampling) is presented in units of kpc at the given distance. The optical radius $r_{\rm 25}$ is from \citet{devaucouleurs1991third}. \\
    Distance references: (a) \citet{tully2013cosmicflows} (b) \citet{anand2021distances} (c) \citet{kourkchi2017galaxy} (d) \citet{tully2016cosmicflows} (e) \citet{tully2009extragalactic}\\
    References for position angle (P.A.), inclination (Incl.), \Mstar\, and SFR: (1) \citet{koribalski2018local} (2) \citet{lang2020phangs} (3) \citet{mateo1998dwarf} (4) \citet{hunter2012littlethings} (5) \citet{leroy2019z} (6) \citet{madden2013overview}\\
    $^{\dagger}$This galaxy has an optical PA (of its stellar bar) notably different from the H{\sc i} disk.\\
    $^{*}$The MW extinction value is adopted from the \citet{schlafly2011measuring} dust maps (see Section~\ref{sec:photometric_images}).}
    \label{tab:sample}
\end{table*}

\subsubsection{Dust extinction correction}
We correct the dust extinction effect for the extracted emission lines for the subsequent analysis, using the Balmer decrement and its conversion to $E(B-V)$ using the following equation:
\begin{equation}
    E(B-V) = \frac{\log \left( \frac{ H\alpha/H\beta (\text{observed}) }{H\alpha/H\beta (\text{intrinsic})} \right) }{0.4 \times (k (H\beta) - k (H\alpha))}, 
    \label{eq:dust_extinction_corr}
\end{equation}
where H$\alpha$/H$\beta$ (intrinsic) is 2.86 for case B recombination (\citealt{osterbrock1989astrophysics}) at $T_{\rm e}$ = 10,000~K and $n_{\rm e}$ of 100~cm$^{-3}$. The $k$(H$\alpha$) and $k$(H$\beta$) represent the amount of extinction at the wavelengths of H$\alpha$ and H$\beta$, respectively, assuming $R(V)$ to be 3.1 based on the Milky Way (MW) extinction curve (\citealt{fitzpatrick1999correcting}). The $k(\lambda)$ is derived from $k(\lambda) = A (\lambda)/E(B-V)$. 

\subsection{Ancillary data}
Together with the optical IFU data, we collect archival ancillary multi-wavelength data obtained from various surveys for each galaxy. In the subsequent sections, we describe the data collected specifically for this study.

\subsubsection{H{\sc i} and CO data}
\label{sec:radio_data}
We obtain H{\sc i} 21~cm and carbon monoxide (CO) radio observation data from various surveys or individual galaxy studies to explore the gas distribution of sample galaxies. The names of the surveys or telescopes are listed in Table~\ref{tab:hi_co_source}. The H{\sc i} 21cm hyperfine transition observation enables us to trace the neutral hydrogen in galaxies. The observations for CO (1-0) or (2-1) rotational transition lines are commonly used to trace molecular hydrogen (H$_2$) in galaxies, as CO molecules can be easily excited in cold molecular clouds.

The H{\sc i} 21~cm data is retrieved from surveys such as The H{\sc i} Nearby Galaxy Survey (THINGS\footnote{\url{https://www2.mpia-hd.mpg.de/THINGS/Overview.html}}; \citealt{walter2008things}), VLA Imaging of Virgo in Atomic Gas (VIVA\footnote{\url{http://www.astro.yale.edu/viva/}}; \citealt{chung2009vla}) survey, Local Irregulars That Trace Luminosity Extremes, The H{\sc i} Nearby Galaxy Survey (LITTLE THINGS\footnote{\url{https://science.nrao.edu/science/surveys/littlethings}}; \citealt{hunter2012littlethings}), Local Volume H{\sc i} Survey (LVHIS\footnote{\url{www.atnf.csiro.au/research/LVHIS}}; \citealt{koribalski2018local}), The Wide-field ASKAP L-band Legacy All-sky Blind surveY (WALLABY\footnote{\url{https://wallaby-survey.org/}}; \citealt{koribalski2020wallaby}) and individual galaxy observation (NGC~6822; \citealt{deblok2000evidence}). We use the flux density map (moment 0 map) derived from natural-weighted H{\sc i} data cubes extracted by the survey teams. For NGC~625, NGC~1705 and NGC~5253, the spatial resolution of the natural-weighted cube is larger than one of the final resolutions of this study (25\arcsec\ at SPIRE~350~$\mu$m; see Section \ref{sec:2.3}) and thus, we instead use robust-weighted data cube. For NGC~1566 and NGC~6822, we extract moment 0 maps from the data cubes using \texttt{3D-Barolo} (\citealt{diteodoro20153dbarolo}) only allowing sources having signal-to-noise ratio (S/N) $>$ 3 (see Appendices of \citealt{verheijen} and \citealt{lelli2014triggering} for noise map calculation).

We search for either CO~(1\--0) or CO~(2\--1) data for individual galaxies based on their availability. For five of the sample galaxies, the CO~(2\--1) images are provided by Physics at High Angular resolution in Nearby Galaxies (PHANGS\footnote{\url{https://sites.google.com/view/phangs/home}}) team as part of the PHANGS\--ALMA collaboration, an Atacama Large Millimeter/sub-millimetre Array (ALMA) large program (\citealt{leroy2021phangsalmadata}; \citealt{leroy2021phangsalma}). This project has extensively mapped the CO J 2$\rightarrow$1 transition at a scale of approximately 100~pc using the ALMA main array (12m), Morita Atacama Compact Array (ACA) (7m), and the Total Power (TP) for 90 nearby galaxies. For NGC~5236 CO (2-1) data, the initial data collection has been made by separate ALMA projects (Project code: 2015.1.00121.S and 2016.1.00386.S, PI: Sakamoto) and run by PHANGS-ALMA pipeline. We obtain the CO~(2-1) moment 0 maps of the five galaxies from the PHANGS data archive\footnote{\url{https://www.canfar.net/storage/list/phangs/RELEASES/PHANGS-ALMA/}}. Another five galaxies in our sample have CO data in the ALMA archive\footnote{\url{https://almascience.nrao.edu/aq/}}. This includes three galaxies, NGC~625, NGC~1705, and NGC~5253, from \citet{hunt2023gas} from ALMA 12m and ACA 7m CO~(1-0) observations (Program code: 2018.1.00219.S, PI: Hunt) and NGC~6822 from ALMA 7m (Program code: 2019.2.00110.S, PI: Kohno; Program code: 2021.1.00330.S, PI: Tosaki). The CO~(1-0) data cubes for these four galaxies were obtained in a primary beam-corrected format initially. We extract the moment 0 maps from the cubes without primary beam correction done using \texttt{3D-Barolo}, ensuring a consistent and well-defined noise level across each channel map. Following the moment 0 map extraction, we subsequently implemented the primary beam correction on these maps. An exception is that the NGC~7793 CO (2-1) moment 0 map has been made separately using the ALMA 12m data cube observed by the ALMA-LEGUS survey (\citealt{grasha2018connecting}; \citealt{finn2024alma}) (Program code: 2015.1.00782.S, PI: Johnson).

For Sextans~A, which has the lowest metallicity (\logOH\ of $\sim$~7.5), there have been several attempts to map CO emissions using ALMA. Unfortunately, robust observations of CO emissions in this galaxy using the interferometer have not yet been successful. For molecular gas information of this galaxy, we find the mass sensitivity limit from the observation implemented on two main star-forming regions of Sextans~A using ALMA (Program code: 2018.1.01783.S; PI: Meyer). We obtain the rms maps using the \texttt{3D-Barolo} for the two native data cubes and multiply by three (i.e., 3~S/N) to construct a {\it CO(1-0) emission sensitivity map}.  

The observation of NGC~6822's H{\sc i} and CO data cubes are affected by the presence of diffuse MW cirrus emission features, which pose significant challenges to the accurate assessment of NGC~6822's H{\sc i} and CO emissions within specific velocity ranges. Specifically, the H{\sc i} emission encounters severe contamination when the systemic velocity lies between -14~\kms and -3~\kms\ (as shown in \citealt{deblok2006stellar}; \citealt{namumba2017hi}; \citealt{park2022gas}), while the CO emission faces similar challenges between 4~\kms\ and 10~\kms\ (\citealt{gratier2010molecular}). To address this issue and minimize the interference caused by the MW cirrus feature, we extract moment 0 maps solely from channels with systemic velocities below -14~\kms\ for H{\sc i} and below 4~\kms\ for CO data. Notably, this selected velocity range does not impact our spatially resolved analysis, given that the central velocity of NGC~6822, where most H II regions are situated, remains near -55~\kms.

\begin{table*}
    \centering
    \begin{tabular}{c|c|c|c|c}
        \hline
        \hline
        Name & H{\sc i} source & & CO source & \\
         & Survey (telescope) & Beam (major $\times$ minor) & Survey (telescope) & Beam (major $\times$ minor) \\
        \hline
        NGC~625 & LVHIS (ATCA)$^{1}$ & 25\arcsec $\times$ 25\arcsec & \--- (ALMA 12m)$^{8}$ & 2.08\arcsec $\times$ 1.49\arcsec \\
        NGC~1512 & LVHIS (ATCA)$^{2}$ & 15\arcsec $\times$ 15\arcsec & PHANGS (ALMA 12m+7m+TP)$^{9}$ & 1.03\arcsec $\times$ 1.03\arcsec \\
        NGC~1566 & WALLABY (ASKAP)$^{3}$ & 42\arcsec $\times$ 35\arcsec & PHANGS (ALMA 12m+7m+TP)$^{9}$ & 1.25\arcsec $\times$ 1.25\arcsec \\
        NGC~1705 & LVHIS (ATCA)$^{1}$ & 25\arcsec $\times$ 25\arcsec & \--- (ALMA 12m)$^{8}$ & 2.49\arcsec $\times$ 1.82\arcsec \\
        NGC~3521 & THINGS (VLA)$^{4}$ & 14.14\arcsec $\times$ 11.15\arcsec & PHANGS (ALMA 12m+7m+TP)$^{9}$ & 1.33\arcsec $\times$ 1.33\arcsec \\
        NGC~4536 & VIVA (VLA)$^{5}$ & 18.04\arcsec $\times$ 16.18\arcsec & PHANGS (ALMA 12m+7m+TP)$^{9}$ & 1.48\arcsec $\times$ 1.48\arcsec \\
        NGC~5236 & THINGS (VLA)$^{4}$ & 15.16\arcsec $\times$ 10.40\arcsec & PHANGS (ALMA 12m+7m+TP)$^{9}$ & 2.14\arcsec $\times$ 2.14\arcsec \\
        NGC~5253 & LVHIS (ATCA)$^{1}$ & 25\arcsec $\times$ 25\arcsec & \--- (ALMA 12m)$^{8}$ & 2.68\arcsec $\times$ 2.41\arcsec \\ 
        NGC~6822 & \--- (ATCA)$^{6}$ & 42.40\arcsec $\times$ 12\arcsec & \--- (ALMA 7m) & (upper) 17.14\arcsec $\times$ 9.55\arcsec \\
         & & & & (middle) 17.11\arcsec $\times$ 8.96 \\
         & & & & (lower) 15.57\arcsec $\times$ 9.33\arcsec \\
        NGC~7793 & THINGS (VLA)$^{4}$ & 15.60\arcsec $\times$ 10.85\arcsec & \--- (ALMA 12m)$^{10}$ & 0.724\arcsec $\times$ 0.724\arcsec \\
        Sextans~A & LITTLE THINGS (VLA)$^{7}$ & 7.6\arcsec $\times$ 6.5\arcsec & \--- & \--- \\
        \hline
        \hline
    \end{tabular}
    \caption{Radio data (H{\sc i}~21~cm and CO) references (1) \citet{koribalski2018local} (2) \citet{koribalski2009gas} (3) \citet{elagali2019wallaby} (4) \citet{walter2008things} (5) \citet{chung2009vla} (6) \citet{deblok2000evidence} (7) \citet{hunter2012littlethings} (8) \citet{hunt2023gas} (9) \citet{leroy2021phangsalma} (10) \citet{grasha2018connecting}}
    \label{tab:hi_co_source}
\end{table*}

\subsubsection{UV, optical, and IR photometric images}
\label{sec:photometric_images}
Most of the photometric UV, optical, and IR data for individual galaxies are retrieved from NED\footnote{\url{https://ned.ipac.caltech.edu/}}. The far-UV and near-UV data for all sample galaxies are obtained as part of the Galaxy Evolution Explorer (GALEX) Nearby Galaxy Survey (NGS; \citealt{gildepaz2007galex}). For the optical wavelength photometric data, we use images taken by various telescopes depending on the availability, including Cerro Tololo Inter-American Observatory (CTIO) 0.9m/1.0m/1.5m, the Kitt Peak National Observatory (KPNO) 2.1m, and the Sloan Digital Sky Survey (SDSS) 2.5m telescopes. We use near-IR and mid-IR images observed with the Two Micron All Sky Survey (2MASS), \textit{Spitzer}\--IRAC/MIPS cameras at wavelengths of 3.5, 4.5, 5.7, 7.8, and 23.7\,$\mu$m, and/or the Wide-field Infrared Survey Explorer (WISE) at wavelengths of 3.4, 4.6, 11.6, and 22.1\,$\mu$m. For the far-IR images, we use data from \textit{Herschel}\--PACS  at wavelengths of 70, 100, and 160$\mu$m, as well as from \textit{Herschel}\--SPIRE at wavelengths of 250 and 350$\mu$m. In Table~\ref{tab:ancillary}, we provide a comprehensive list of the instruments, band names, effective wavelengths, and PSF/beam sizes. In addition, a list of the ancillary data of each sample and their references are available in Table~\ref{tab:ancillary_ref}. All photometric data are extinction corrected for foreground MW dust using dust reddening maps\footnote{\url{https://irsa.ipac.caltech.edu/applications/DUST/}} from \citet{schlafly2011measuring} and assuming the MW extinction curve of \citet{fitzpatrick2019analysis}.

\subsection{Image processing}
\label{sec:2.3}
For a region-by-region analysis, we ensure that all multi-wavelength images have the same spatial resolution as the coarsest image, which in this study is the Herschel-SPIRE~350~$\mu$m image with a resolution of approximately 25\arcsec for the majority of our sample galaxies (except for NGC~1566, NGC~6822, and Sextans~A; see Section~\ref{sec:2.3.3}). We note that the SPIRE~500~$\mu$m images were not considered in this study. The SPIRE~500~$\mu$m band is sensitive to the massive, cold dust components. However, excluding these images still allows us to reliably estimate the interstellar \Mdust\ of a system, with < 30\% of error from the dust mass estimated using far-IR bands including 500~$\mu$m (\citealt{aniano2012modeling}). Additionally, it is important to note that the SPIRE~500~$\mu$m image has a considerably coarse resolution of 37.5\arcsec, thus excluding it allows us to obtain more pixels with reliable \Mdust\ at the same time. 

In the following sections, we describe the image processing steps applied to all data being used in this work for region-by-region analysis. The units of the images at wavelengths from far-UV to far-IR (photometry images hereafter) are converted to Jy pix$^{-1}$ at the end of the image process (i.e., after image convolution and resampling are completed) to match with the unit required to run the SED modelling code \texttt{MAGPHYS} (see Section~\ref{sec:magphys}).

\subsubsection{Sky background subtraction}
The photometry images have been sky background subtracted. While many survey images are already sky background subtracted through their data reduction pipelines, some may not have undergone this process. For such images, we build sky background maps using an IDL tool `sky'\footnote{\url{https://idlastro.gsfc.nasa.gov/ftp/pro/idlphot/sky.pro}}. We apply a 3-$\sigma$ clipping method to remove sources from the image and derive a median value of the background levels of a given box size. Subsequently, the measured sky background images are subtracted from the original images.

\subsubsection{Foreground star removal}
It is crucial to address the impact of bright foreground stars on galaxy images, particularly for low-luminosity, dwarf galaxies. These stars can introduce significant contamination and affect measurements of intrinsic \Mstar\ or SFR, making it challenging to get reliable SED modelling, especially for galaxies located along the line-of-sight of the MW disk. To mitigate this effect, similar to the approach employed by \citet{clark2018dustpedia} and \citet{verstocken2020high} carried out for the DustPedia project (\citealt{davies2017dustpedia}), we utilize the \texttt{PTS-7/8}\footnote{\url{https://github.com/SKIRT/PTS}} (\citealt{verstocken2020high}) software, the Python toolkit associated to the SKIRT radiative transfer code (\citealt{camps2015skirt}; \citealt{camps2020skirt}). We remove the foreground stars from the GALEX far-UV up to {\it Spitzer} 24$\mu$m images. The \texttt{PTS-7/8} software retrieves the source catalogue from the 2MASS All-Sky Catalog of Point Sources (\citealt{cutri20032mass}), identifies point sources in the input image, and performs interpolation using the surrounding pixels within a region defined by the FWHM of each point source. The contaminated pixels are then replaced through interpolation. 

Given NGC 6822's wide span across the sky with its central 12\arcmin-wide stellar bar and its closeness to the MW plane, we include an additional step in the foreground source removal process. NGC~6822 is very close to us with a distance of $\sim$~0.50~Mpc, such that the stars within the galaxy are also in the 2MASS Catalog of Point Sources with severe contamination from MW foreground stars. To distinguish stars belonging to NGC~6822 from the MW foreground stars, we combine two source catalogues provided by \citet{hirschauer2020dusty} and Gaia DR3 catalogue (\citealt{gaia2016gaia}; \citealt{gaia2023gaiadr3}). We initially run the \texttt{PTS-7/8} through the far-UV to mid-IR multi-wavelength images of NGC~6822, to detect and compile a catalogue of all identified point sources for subsequent masking purposes. To distinguish Milky Way/main sequence (MW/MS) stars within NGC~6822, we perform a cross-match of coordinates between the categorized MW/MS sources (from colour-magnitude diagram by \citealt{hirschauer2020dusty}) and the Gaia DR3 source. Allowing a maximum separation of 1\arcsec between corresponding sources in both catalogues enables us to classify MW stars located along the line of sight of NGC~6822, effectively isolating MS stars within NGC~6822. Subsequently, we remove the stars identified as likely belonging to NGC~6822 from the source catalogue found earlier by \texttt{PTS-7/8}, preserving solely the MW point sources for eventual removal from the input images.

\subsubsection{Image convolution and resampling}
\label{sec:2.3.3}
As the next step, we perform image convolution for all images including TYPHOON emission line maps, the photometric images, and the CO and H{\sc i} flux density maps to match with the resolution of SPIRE~350$\mu$m band (25\arcsec). Since each instrument has its unique point spread function (PSF), the choice of individual kernels for the convolution process depends on the characteristics of both the initial PSF and the desired PSF. To facilitate the convolution process, we utilize a set of kernels provided by \citet{aniano2011}\footnote{\url{https://www.astro.princeton.edu/~draine/Kernels.html}}, along with the corresponding IDL code. This enables us to readily implement convolution on the given images to achieve the desired resolution. 

For the cases of NGC~1566 and NGC~6822, the final beam size is set to larger than the other galaxies as the coarsest resolution among the used multi-wavelength data is not SPIRE~350~$\mu$m, but H{\sc i} data (see Table~\ref{tab:hi_co_source} and Table~\ref{tab:ancillary}). Considering the major beam sizes of NGC~1566 and NGC~6822 H{\sc i} data are 42\arcsec and 42.4\arcsec, respectively, we set 43\arcsec as their final resolution. In the convolution process, we use the closest desired resolution of 41\arcsec using kernels provided by {\citet{aniano2011}}.

Once the convolution process is complete, we resample the images using \texttt{SWARP}\footnote{\url{https://www.astromatic.net/software/swarp/}} (\citealt{bertin2002terapix}), ensuring that the pixel scale matches the PSF of the coarsest image, which has a spatial resolution of 25\arcsec, or 43\arcsec for NGC~1566. Resampling the final pixel to be comparable with the PSF size of the SPIRE~350~$\mu$m ensures that our regions can be considered independent.

For the cases of NGC~6822 and Sextans~A, as an exception, we resample the convolved multi-wavelength images including TYPHOON data into the physical resolution of 200~pc per pixel (i.e.,  82.5\arcsec and 29.7\arcsec, respectively). This is to ensure reliable dust masses, which occur for regions with log \Mdust $>$ 3 (see Figure 10 of \citealt{smith2018panchromatic}). This threshold is exceeded for the regions in NGC~6822 and Sextans~A considered in this work.

\subsubsection{Photometric uncertainty measurement}
The photometric uncertainty associated with each pixel in the given image is determined by the standard combination of the Poisson error, the scatter in the sky values, and the uncertainty in the mean sky. The calculation of the uncertainty is expressed as follows: 
\begin{equation}
    \sigma = \sqrt{f(\lambda) + A(ap) \times sky^2 + \frac{A(ap)^2 \times sky^2}{A(sky)}},
    \label{eq:uncertainty}
\end{equation}
where f($\lambda$) indicates the number of photons received at a given wavelength (i.e., Poisson error), $A(ap)$ denotes the area of the galaxy's aperture measured in pixels squared, $sky$ corresponds to the standard deviation of the sky values per native pixel, and $A(sky)$ represents the area of the sky apertures in native pixels squared. $sky$ and $A(sky)$ values are measured using {\sc sky.pro}, an IDL photometry tool. The unit conversion, image convolution, and resampling processes are also applied to these uncertainty maps as implemented for the flux density maps.

\section{Method}
\label{sec:physicalparam}
In this section, we describe the method we employ to measure the physical parameters including the \Mdust\ using a SED modelling tool, gas-phase metallicity, and gas masses (H{\sc i} and H$_{2}$ components) of each region in individual galaxies.

\subsection{SED modelling: MAGPHYS}
\label{sec:magphys}
We use the \texttt{MAGPHYS} (Multi-wavelength Analysis of Galaxy Physical Properties) framework (\citealt{dacunha2008}; \citealt{battisti2019}) to perform spectral energy distribution (SED) modelling and derive various physical properties in a spatially resolved manner, including the \Mstar, SFR, and \Mdust. \texttt{MAGPHYS} is a powerful tool designed to analyze and interpret the observed UV to the sub-mm luminosity of a given sample by constructing optical and infrared emission libraries that account for both stellar and dust components. The code builds optical libraries assuming the Chabrier initial mass function (IMF; \citealt{chabrier2003}), the stellar population synthesis model from \citet{bruzual2003stellar}, and metallicity range from 0.2 $<$ $Z$/\Zsol $<$ 2 at the given redshift. The star formation history (SFH) is modelled with a combination of an underlying continuous model and randomly superimposed starbursts to approximate more realistic SFHs. The dust attenuation effect is accounted for by considering two components: the birth cloud (age less than 10 Myr) and the surrounding ISM (age more than 10 Myr), using the prescription from \citet{charlot2000simple}. To construct the infrared libraries, \texttt{MAGPHYS} assumes that the attenuated stellar emission by dust is re-emitted at mid-IR through far-IR wavelengths, thereby maintaining energy balance. The infrared libraries are built with a sum of three components: Polycyclic aromatic hydrocarbons (PAHs), warm dust components ($T_{\rm dust}$ of 30\---70 K), and cold dust components ($T_{\rm dust}$ of 15\---25 K). The code then compares the modelled SED with the observed SED, calculates $\chi^2$, and finds the best-fit combination of models. The probability distribution functions (PDFs) are calculated for individual physical parameters based on the likelihood distribution from the Bayesian approach.

\begin{figure*}
    \centering
    \includegraphics[scale=0.44]{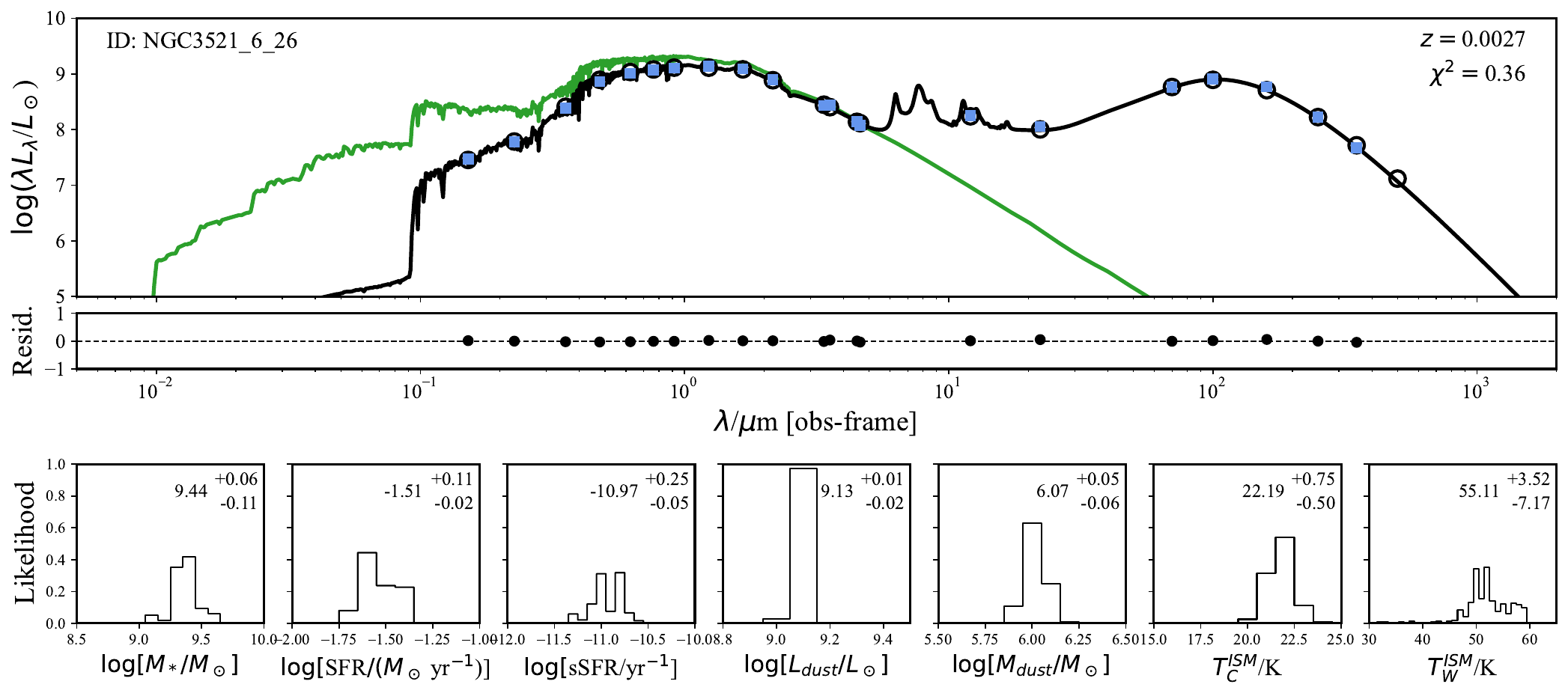}
    \caption{Example of SED fit results derived by \texttt{MAGPHYS} for a spaxel in NGC~3521. The top half panel presents the best-fit attenuated SED model (black solid line) and the unattenuated SED model (green solid line) for the given pixel coordinate, along with measured photometry (blue squares) and their errors. The spectroscopic redshift and $\chi^2$ value are shown in the top right corner. The lower portion displays the residuals between the model and observed luminosities. The bottom half panels show the likelihood distributions of key parameters: \Mstar, SFR, specific star formation rate (sSFR), dust luminosity ($L_{\rm dust}$), \Mdust, cold dust temperature in the ambient ISM ($T_{\rm C}^{\rm ISM}$), and warm dust temperature in the ISM ($T_{\rm W}^{\rm ISM}$). The numbers in the top right of each panel represent the median values, while the 16\% (`-') and 84\% (`+') symbols indicate the standard deviation.}
    \label{fig:example_magphys}
\end{figure*}

Unlike conventional usage where integrated fluxes of entire galaxies are considered, we focus on characterizing properties at a smaller scale, which would likely be smaller compared to integrated measurements. To take this into account, we adjust the posterior range of several physical parameters in the \texttt{MAGPHYS} code. Specifically, we modify the code to allow lower \Mstar\ (5 $<$ log~\Mstar/\Msol $<$ 11), SFR (-5 $<$ log~SFR/\Msol~yr$^{-1}$ $<$ 1.5), dust luminosity (5 $<$ log~$L_{\rm dust}$/$L_{\odot}$ $<$ 10), and \Mdust\ (2 $<$ log~\Mdust/\Msol $<$ 7). Figure~\ref{fig:example_magphys} shows an example of the SED fit result derived by \texttt{MAGPHYS} for a spaxel of NGC~3521.

We exclude pixels with calculated $\chi^2$ values larger than $\mu$(${\chi^2}$) + 3~$\times$~$\sigma$(${\chi^2}$), where $\mu$(${\chi^2}$) and $\sigma$(${\chi^2}$) are the median and standard deviation, of the obtained $\chi^2$ distributions of individual galaxies. Additionally, we remove pixels with \Mdust\ probability distribution that spreads out over a wide range, as this indicates unreliable median \Mdust\ values. To achieve this, we calculate the ratio between the median \Mdust\ and the standard deviation (averaging the 16\% and 84\% percentiles) and exclude pixels with a ratio smaller than 1.

\subsection{Gas-phase metallicity: Scal}
\label{sec:scal}
Oxygen is the most abundant element in the universe following hydrogen and helium and does not undergo significant depletion onto dust grains (\citealt{draine2011physics}). Thus, the relative abundance of oxygen to hydrogen in \textit{gas-phase}, commonly represented as \logOH, can serve as a proxy for the overall metal abundance in the ISM (\citealt{kewley2019understanding}). Oxygen abundances are most accurately obtained using temperature-sensitive auroral lines that rely on direct electron temperature measurements. However, in low-excitation, oxygen-rich regions, these lines are often too weak to be detected. Consequently, numerous studies have attempted to empirically derive the gas phase metallicity through various combinations of strong emission lines (i.e., `strong-line' methods), such as R$_{23}$ (\citealt{pagel1979on}), N2H$\alpha$ (\citealt{storchi-bergmann1994ultraviolet}), and Scal (\citealt{pilyugin2016new}). Theoretical approaches (e.g., N2O2 \--- \citealt{kewley2002using}; N2S2H$\alpha$ \--- \citealt{dopita2016}) and combinations of empirical and theoretical methods (e.g., O3N2 \--- \citealt{pettini2004abundance}; N2 \---\citealt{denicolo2002new}) have also been developed to determine gas-phase metallicity. 

For our analysis, we choose the Scal metallicity diagnostic (\citealt{pilyugin2016new}). Among the many popular metallicity tracers, we exclude diagnostics that rely on [O~II] $\lambda$3726, $\lambda$3729 emission lines. These lines are located at the blue wavelength end of TYPHOON spectra, where the sensitivity is notably lower relative to the red end. The Scal calibration has been shown to agree with $T_{\rm e}$-based abundances within $\sim$~0.1~dex over a wide metallicity range (7.0 $<$ \logOH $<$ 8.6; \citealt{pilyugin2016new}). Also, this calibration shows relatively low sensitivity towards variation in gas pressure or ionization parameters. Moreover, as described in \citet{devis2017using}, this diagnostic is one of the most reliable metallicity calibrations for low-metallicity sources.

The Scal calibration uses the following three ratios:
{\hspace{0.2cm}
\begin{align*}
    &\rm N_2 = ([N~II]\lambda~6548 + 6583)/H\beta, \\
    &\rm S_2 = ([S~II]\lambda~6716 + 6731)/H\beta, \\
    &\rm R_3 = ([O~III]\lambda~4959 + 5007)/H\beta,
\end{align*}
and divides lower and higher branches defined by log N$_2$. We apply the intrinsic [N~II] and [O~III] intensity ratio fixed to the ratio given by quantum mechanics (\citealt{storey2000theoretical}), as their ratios are independent of physical conditions ([N~II]$\lambda$~6583 / [N~II]$\lambda$~6548 = 3.05 and [O~III]$\lambda$~5007 / [O~III]$\lambda$~4959 = 2.98). The oxygen abundance at the lower branch which has log N$_2$ $\leq -$ 0.6 is described as
\begin{align}
    \begin{split}
        &\rm \logOH_L \\
        &\rm = 8.072~+~0.789~log(R_3/S_2)~+~0.751 logN_2 \\
        &\rm~~~~+~(1.069~-~0.170 log(R_3/S_2)~+~0.022~logN_2) \times logS_2,
    \end{split}
    \label{eq:scal_L}
\end{align}
while the oxygen abundance at the higher branch (log N$_2$ $> -$ 0.6) can be calculated as
\begin{align}
    \begin{split}
        &\rm \logOH_H \\
        &\rm = 8.424~+~0.030~log(R_3/S_2)~+~0.751~logN_2 \\
        &\rm ~~~~+~(-0.349~+~0.182~log(R_3/S_2)~+~0.508~logN_2) \times logS_2.
    \end{split}
    \label{eq:scal_H}
\end{align}}
We note that the lines used here are dust extinction corrected using Equation~\ref{eq:dust_extinction_corr}. Throughout the subsequent sections, we use the terms gas-phase metallicity (\Zgas) and \logOH\ interchangeably.

However, it is well known that the different metallicity diagnostic provides large discrepancies in absolute oxygen abundances from one another diagnostic, which can vary up to 0.7~dex (e.g., \citealt{kewley2008metallicity}; \citealt{poetrodjojo2019effects}; \citealt{groves2023phangs}). Although this is out of the scope of this study, we present our main analysis with different metallicity diagnostics in Appendix~\ref{app:diff_calibrations}.

\subsection{Gas mass}
\label{sec:mgas}
H{\sc i} 21~cm emission lines are often assumed to be optically thin, allowing us to directly convert the observed H{\sc i} 21~cm flux density to the H{\sc i} gas mass (\MHI) using the following equation (\citealt{roberts1962neutral}):
\begin{equation}
    \rm \left(\frac{\rm M_{HI}}{\rm M_{\odot}}\right) = 2.36 \times 10^{5} \left(\frac{D}{Mpc} \right)^{2} \left(\frac{S^{HI}_{int}}{Jy\,km s^{-1}}\right) (1+z)^{-2},
    \label{eq:himass}
\end{equation}
where $D$ is the distance of the galaxy, $S\rm ^{HI}_{int}$ is the integrated H{\sc i} flux, and $z$ is the redshift. We use the redshift to calculate the luminosity distance, to maintain consistency when comparing and combining H{\sc i} gas mass measurements with other physical properties (\Mstar\ and \Mdust) derived from \texttt{MAGPHYS} SED fit. We note that the distance terms cancel out in the DGR or other parameters regarding mass ratios of ISM tracers.

Due to the lack of a dipole moment in molecular hydrogen, a direct measurement of H$_2$ molecular gas mass is challenging. Therefore, many studies resort to using $^{12}$CO~(1-0) observations as a proxy, as CO is the next most abundant molecule after H$_2$ and tends to trace similar regions. $^{12}$CO~(2-1) rotational transition is widely used alternatively thanks to its relatively cheaper observations ($\sim$~2\---4 times faster than mapping $^{12}$CO~(1-0) to the same mass surface density, sensitivity, and angular resolution; \citealt{leroy2021phangsalma}). With the $^{12}$CO~(2-1) observations, we convert the measured $^{12}$CO~(2-1) intensity to $^{12}$CO~(1-0) first using the intensity ratio between two lines ($R_{\rm 21}$), and then to H$_{2}$ surface density using the conversion factor ($\alpha_{\rm CO}$) using following empirical relations.
{\hspace{0.3cm}
\begin{equation}
    I_{\rm ^{12}CO(1-0)} = R_{\rm 21}^{-1} \times I_{\rm ^{12}CO(2-1)},
    \label{eq:co21toco10}
\end{equation}
where $R_{21}$ is assumed as 0.63 (\citealt{leroy2021phangsalma}; \citealt{denbrok2021new}). \citet{denbrok2021new} found the luminosity-weighted mean $R_{\rm 21} = 0.63 \pm 0.09$, which is included in the uncertainty measurement.

The derived $^{12}$CO (1-0) is converted into $\Sigma_{\rm mol}$ ($\Sigma_{H_2}$ $\times$ 1.36 to correct Helium abundance) under the assumption of optically thick $^{12}$CO (1-0) emission.
\begin{equation}
    \Sigma_{\rm mol} = \left(\frac{\alpha_{CO}}{M_{\odot}pc^{-2}}\right)~\left(\frac{I_{^{12}CO(1-0)}}{K~km~s^{-1}}\right)~\mathrm{cos}~{(i)}~(1+z)^{-1},
    \label{eq:co10tomh2}
\end{equation}}
where {\it i} is the inclination of the galaxy. For NGC~625, NGC~1705, NGC~5253, and NGC~6822, as we use $^{12}$CO~(1-0) intensities, we directly convert the observed $^{12}$CO~(1-0) intensity to the molecular gas surface density ($\Sigma_{mol}$) without applying $R_{\rm 21}$. The molecular gas mass (\Mmol) is calculated from the estimated $\Sigma_{\rm mol}$.

There have been many studies investigating its dependency on the metallicity of galaxies or regions, finding that it tends to increase as the metallicity decreases (e.g., \citealt{amorin2016}, \citealt{chiang2023kpc}). \citet{amorin2016} explored this relationship in 21 blue compact dwarf galaxies and proposed a metallicity-dependent conversion factor described by the equation:
\begin{equation}
    {\rm log}\,(\alpha_{\rm CO, Z}) = 0.68 - 1.45 \times ({\rm \logOH} - 8.7).
\end{equation}
This equation highlights a possible underestimation of \Mmol\ when using the constant \alphacomw\ value, which is measured as 4.35 \Msol$\rm pc^{-2}$ (K~\kms)$^{-1}$ (\citealt{strong1996gradient}). However, the accuracy of the $\alpha_{\rm CO}$ is still under debate (\citealt{bolatto2013}) thus we mainly
focus on the \alphacomw. which is measured from the MW for
further analysis. (\citealt{bolatto2013}). We note that these factors are helium abundance corrected (36~\%). The total gas mass of the region is measured by summing the \MHI\ and \Mmol\ as \Mgas\ = 1.36 $\times$ \MHI\ + \Mmol. The helium contribution is also corrected for the H{\sc i} gas mass by multiplying a factor of 1.36. 

\subsection{H~II region selection: BPT diagram}
\label{sec:bpt}
To ensure accurate measurements of the metallicity in specific regions, we exclusively consider star-forming regions, as our metallicity diagnostics in Section~\ref{sec:scal} rely on physical conditions that are typical for star-forming H~II regions. Hence, it is crucial to exclude pixels heavily affected by active galactic nuclei (AGN) or shocks. To achieve this, we employ the Baldwin-Phillips-Terlevich (BPT) diagram introduced by \citet{baldwin1981classification}. To ensure the accuracy of our metallicity measurements in Section~\ref{sec:scal}, we adopt the \citet{kauffmann2003host} separation (blue solid line) to focus solely on genuine star-forming regions. The star-forming H~II regions used in our resolved analysis are shown in black dots.

\begin{figure}
    \centering
    \includegraphics[scale=0.54]{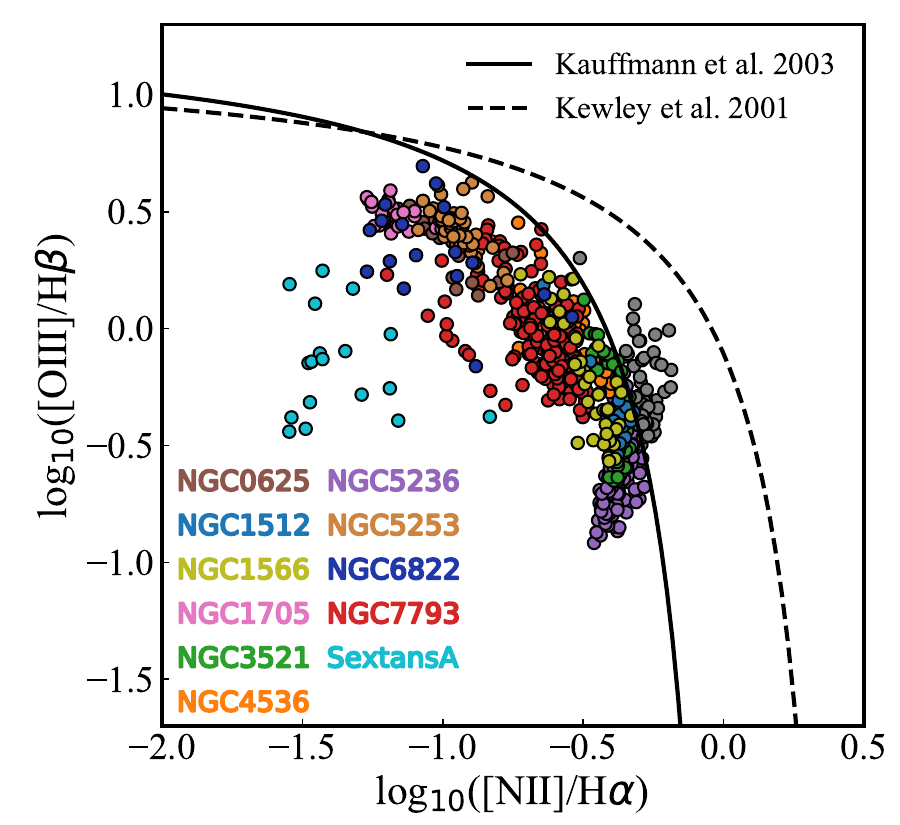}
    \caption{The BPT diagram for all regions in sample galaxies. The filled circles are colour-coded by galaxy. The black dashed line indicates the criteria defined by \citet{kewley2001theoretical}. In this study, we use \citet{kauffmann2003host} criteria to be conservative for star-forming H~II region selection. Grey circles present regions where affected by other sources than star-forming activity.}
    \label{fig:bpt}
\end{figure}

\section{Result}
\label{sec:result}

\subsection{The global DGR\---\Zgas\ relation}
\label{sec:global_dgr_zgas}
As a sanity check on our methodology before examining spatially resolved relationships, we first investigate the relationship between the dust-to-gas mass ratio (DGR) and the gas-phase metallicity (\Zgas) of our sample galaxies on a global scale. To obtain the galaxy-integrated physical parameters such as \Mdust, \Mgas\ (\Matom\ and \Mmol) and \Mstar for individual galaxies, we sum up the measured quantity of each parameter of pixels within the optical radius ($r_{25}$) listed in Table~\ref{tab:sample}.

We measure the global metallicity (with Scal calibration) of the individual galaxies based on integrated emission line ratios of star-forming H~II regions within each galaxy. Specifically, we sum up the emission line fluxes required for the metallicity diagnostic from star-forming H~II regions within $r_{25}$. We then calculate emission line ratios using the metallicity diagnostic equations (Equation~\ref{eq:scal_H}). We highlight that the wide FoV of TYPHOON data allows us to measure reliable \Zgas, encompassing the bulk of SF regions and ISM. The DGR values assuming \alphacoz\ tend to be smaller than those with \alphacomw\, aligning with expectations since \alphacoz\ increases at lower metallicity. However, we observed minimal deviation between the two DGR measurements, with a median discrepancy of less than 0.1 dex.

\begin{table}
    \centering
    \begin{tabular}{c|c|c|c}
        \hline
        \hline
        Name & \multicolumn{1}{c}{\logOH} & \multicolumn{2}{c}{log$_{10}$ DGR} \\
        Calibration & Scal & \alphacomw & \alphacoz\ - Scal \\
        \hline
        NGC~625 & 8.167~$\pm$~0.005 & -2.48~$\pm$~0.04 & -2.51~$\pm$~0.06 \\
        NGC~1512 & 8.567~$\pm$~0.010 & -2.82~$\pm$~0.01 & -2.83~$\pm$~0.01 \\
        NGC~1566 & 8.549~$\pm$~0.005 & -2.53~$\pm$~0.03 & -2.62~$\pm$~0.03 \\
        NGC~1705 & 8.087~$\pm$~0.005 & -2.94~$\pm$~0.12 & -2.96~$\pm$~0.16 \\
        NGC~3521 & 8.573~$\pm$~0.003 & -2.30~$\pm$~0.01 & -2.34~$\pm$~0.01 \\
        NGC~4536 & 8.503~$\pm$~0.006 & -2.32~$\pm$~0.01 & -2.41~$\pm$~0.01 \\
        NGC~5236 & 8.591~$\pm$~0.002 & -2.20~$\pm$~0.01 & -2.36~$\pm$~0.01 \\
        NGC~5253 & 8.199~$\pm$~0.003 & -2.37~$\pm$~0.02 & -2.41~$\pm$~0.02 \\
        NGC~6822 & 8.191~$\pm$~0.012 & -2.87~$\pm$~0.05 & -2.87~$\pm$~0.03 \\
        NGC~7793 & 8.354~$\pm$~0.006 & -2.32~$\pm$~0.01 & -2.42~$\pm$~0.01 \\
        \multirow{2}{*}{Sextans~A} & \multirow{2}{*}{7.536~$\pm$~0.022} & \multirow{2}{*}{$>$ -4.27~$\pm$~0.09} & $>$ -4.33$\pm$~0.09 \\
         & & & ($>$ -4.96)  \\
        \hline
        \hline
    \end{tabular}
    \caption{Measurements of \logOH\ (Scal) and DGR (\alphacomw\ and \alphacoz) within optical radius ($r_{\rm 25}$). The uncertainty is calculated from the error propagation, taking into account each pixel's uncertainty. For Sextans~A, we present the DGR from \Mmol\ limit using \alphacomw, \alphacoz, and [C~II]-based \Mmol\ in parenthesis.}
    \label{tab:globalvalues}
\end{table}

\begin{figure}
    \centering
    \includegraphics[scale=0.55]{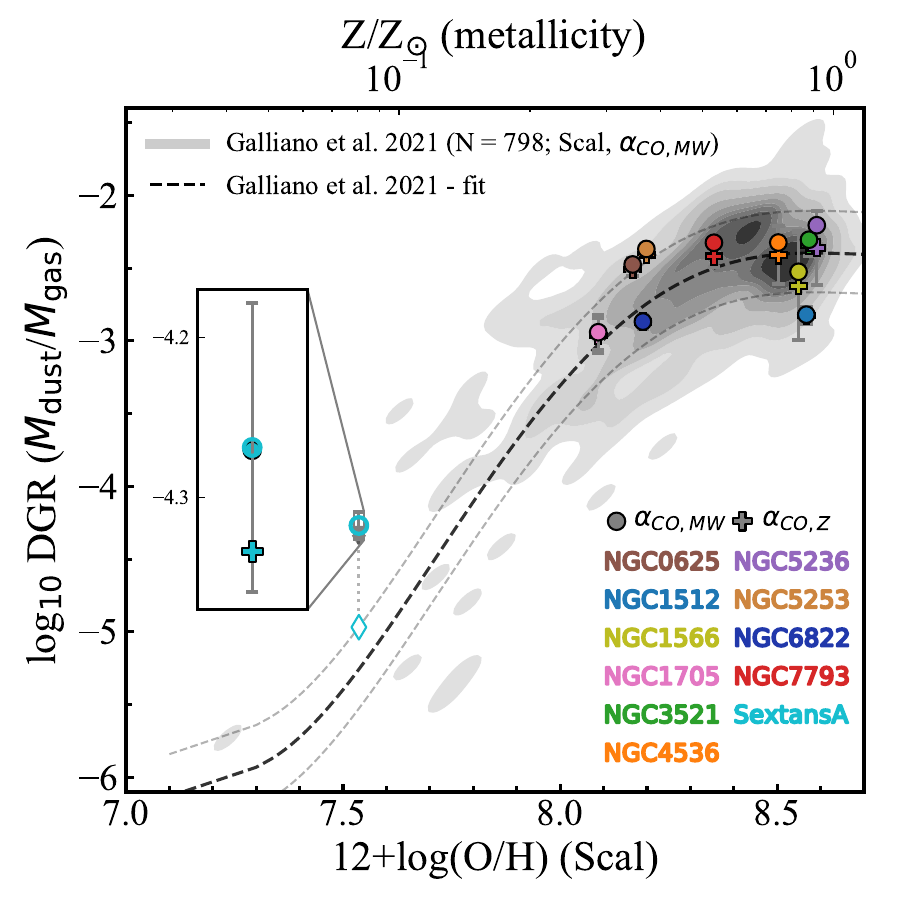}
    \caption{The global DGR\---\Zgas\ relationship. The x-axis shows the oxygen abundance (i.e., \Zgas) derived using the Scal calibration. The DGR values of each galaxy are from \Mmol\ both \alphacomw\ and \alphacoz\ and are shown with the circles and pluses. The top x-axis presents the metallicity in a unit of solar metallicity converted from the oxygen abundance with an assumption of \Zsol\ = 8.69 (\citealt{asplund2009chemical}) and the composition ratio of other chemical elements is the same as the sun. The cyan open circle presents Sextans~A whose \Mgas\ only considers the detected \Matom\ and the mass sensitivity limit from the CO rms map. The cyan open diamond symbol connected to the open circle shows the application of the conservative molecular gas amount from [C~II] emission. The grey background density map shows the literature trend (\citealt{galliano2021nearby}, applicable only for any comparison with DGR \alphacomw). The sub-plot focusing on Sextans~A presents the DGR lower limit from the assumption of \alphacomw\ (filled circle, not visible due to its similar value with DGR value without \Mmol; shown as the open circle) and \alphacoz\ (filled plus) for \Mmol\ using the CO sensitivity map. The black and grey dashed lines represent the fourth polynomial fit from the literature and the 1$\sigma$ uncertainty, respectively. The colour-coded symbols are the same as previous figures.}
    \label{fig:global_dgr_zgas}
\end{figure}

Figure~\ref{fig:global_dgr_zgas} illustrates the relationship between global DGR and \Zgas, showing data points colour-coded by galaxy. Our DGR measurements incorporate \Mmol\ estimations derived from the constant \alphacomw\ (star-shaped points) and the metallicity-dependent \alphacoz\ (circular points). We show the DGR measurement for Sextans~A as an open circle to show an upper limit as only the \Matom\ is considered. Additionally, we present two lower bounds of the DGR in the embedded panel based on the \Mmol\ mass limit derived from its CO~(1\--0) emission sensitivity map (see Section~\ref{sec:radio_data}), by integrating noise values of pixels in the aperture size chosen in this study (i.e., $r_{25}$). The plus symbol shows the DGR lower bound obtained using \alphacoz\, and the filled circle shows that using \alphacomw. We note that the \Mmol\ limit from \alphacomw\ plays a negligible role in DGR measurements, making the lower bound of DGR hardly seen (overlapping with the open circle). The open diamond symbol connected to the circle presents the `conservative' DGR values calculated from the \MHt/\MHI\ of 4 from [C~II] 158$\mu$m measurement in \citet{cigan2021herschel}. We discuss this more in Section~\ref{sec:sextansA}. 

The grey-shaded region shows the density map of data from \citet{galliano2021nearby} of $\sim$800 DustPedia and DGS galaxies. In their work, they derived global DGR (\alphacomw\ $=$ 4.35 \Msol$\rm pc^{-2}$ (K\kms)$^{-1}$) and \Zgas\ (Scal) values. We also show their 4th polynomial fit result and the 1$\sigma$ uncertainty (black and grey dashed lines, respectively; equations 8 and 9 in \citealt{galliano2021nearby}). It is important to note a caveat in this comparison: for some low-metallicity galaxies in their sample, the \Mmol\ values were estimated using a scaling relation between \Matom/\Mstar\ and \Mmol/\Matom\ (see Equation 5 of \citealt{casasola2020ism}) due to the non-detection of CO.

The global DGR\---\Zgas\ relationship for TYPHOON galaxies aligns well with the literature trend, falling within 1$\sigma$ (see circles for a fair comparison in terms of the same calibrations used). Two galaxies, Sextans~A and NGC~1512, are noticeable with their offset from the reference trend. NGC~1512 deviates from the general trend towards lower DGR by $\sim$~0.6~dex. It is attributed to its unique H{\sc i} distribution, with most of the H{\sc i} located at larger radii about $\sim$ 4 times the optical disk (\citealt{koribalski2009gas}; also see Figure~\ref{fig:NGC1512}).

\subsection{The resolved ISM scaling relations}
In the following analysis of the ISM scaling relations, we only show the resolved \Mmol\ or resolved DGR values calculated using the \alphacomw\ to reduce any uncertainty that could be raised by employing \alphacoz. 

\begin{figure}
    \centering
    \includegraphics[scale=0.53]{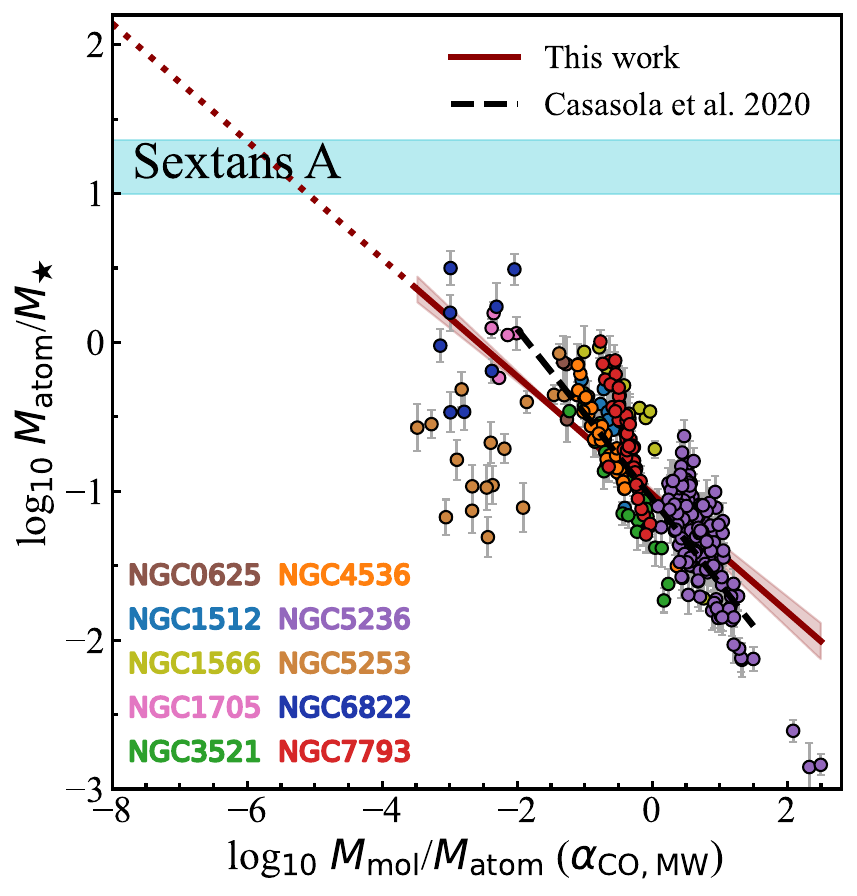}
    \caption{The scaling relation between the atomic gas to stellar mass fraction (\Matom/\Mstar) as a function of the molecular gas fraction (\Mmol/\Matom) in a spatially resolved scale. The \Mmol\ is from the constant \alphacomw. The dark-red solid line presents the linear fit from this study with the extension towards the lower \Mmol/\Matom\ as a dotted line. The black dashed line shows the global measurements from \citet{casasola2020ism}. The cyan stripe indicates where the resolved \Matom/\Mstar\ of Sextans~A belongs. The colour-coded symbols and the reference are the same as previous figures. }
    \label{fig:sr_casasola20}
\end{figure}

{\it {\Mmol/\Matom\---\Matom/\Mstar\ relation}}: We probe the resolved relation between \Matom/\Mstar\ and \Mmol/\Matom, which will give us a hint of how much \Mmol\ would reside in resolved regions in Sextans~A (for the similar approach at global scale, see \citealt{casasola2022resolved}; \citealt{devis2019}; \citealt{galliano2021nearby}). In Figure~\ref{fig:sr_casasola20}, the literature trend from 245 DustPedia late-type galaxies by \citet{casasola2020ism} is shown with a dashed line\footnote{The original relation in their paper is $y = -0.57x - 1.18$. As their scaling relation does not consider helium abundance to \Mmol\ and \Matom, we adjusted their relation by adding a helium abundance correction factor of 1.36 in this figure. This changes the y-intercept of the relation increased by $\sim$~0.133 ($= {\rm log}_{10} 1.36$).}. The dark red solid line shows the best-fit linear relation:
\begin{equation}
    \log_{10} (M_{\rm atom}/M_{\rm \star}) = -0.40 \log_{10} (M_{\rm mol}/M_{\rm atom} -1.02)
\end{equation} with the extrapolated line towards lower \Mmol/\Matom\ region.

Some of the higher mass/metallicity galaxies are moderately consistent with the literature trend (global; \citealt{casasola2020ism}). However, when considering our resolved measurement extending down to \Mmol/\Matom\ of 10$^{-4}$, we find a shallower slope. For dwarf galaxies such as NGC~625, NGC~1705, NGC~5253, and NGC~6822, the correlation between these two parameters is less evident, displaying a broader scatter compared to larger galaxies with higher \Mmol/\Matom\ values. We show the \Matom/\Mstar\ range for regions within Sextans~A, denoted by a cyan stripe. The \Mmol/\Matom\ range predicted from the linear fit extrapolation corresponds to $-6$$\sim$$-5$ in the logarithmic unit, implying these regions are highly dominated by atomic gas.

\begin{figure}
    \centering
    \includegraphics[scale=0.53]{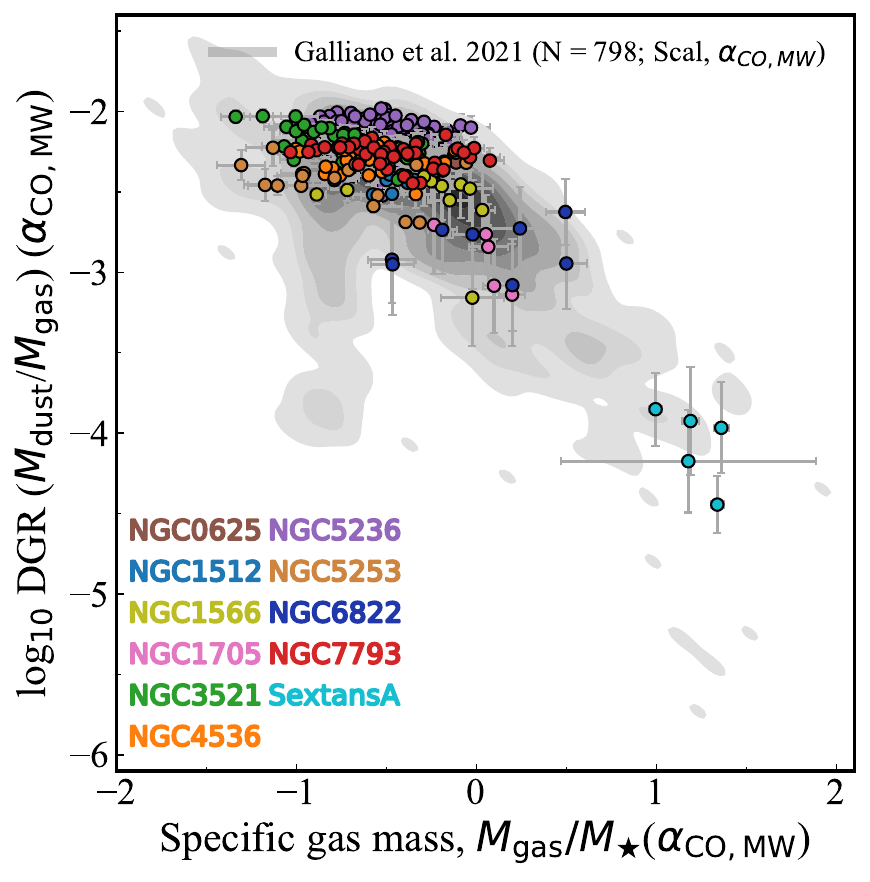}
    \caption{The scaling relation between the DGR as a function of the specific gas mass (\Mgas/\Mstar) in a spatially resolved scale. The background density map shows DustPedia+DGS galaxies studied by \citet{galliano2021nearby}. The colour-coded symbols are the same as previous figures. }
    \label{fig:specific_gas_mass_DGR}
\end{figure}

{\it DGR\---\Mgas/\Mstar\ relation}: In Figure~\ref{fig:specific_gas_mass_DGR}, we show how the DGR of regions in our sample galaxies varies as a function of the specific gas mass (=\Mgas/\Mstar), which can be a tracer of galaxy evolution in the way of more gas in the region being consumed to form stars, decreasing the specific gas mass. We note that the different assumption of $\alpha_{\rm CO}$ for DGR values shows only a minor difference, which is hinted in the global DGR\---\Zgas\ relationship in Figure~\ref{fig:global_dgr_zgas}. As applied to the galaxy-integrated measurement, for Sextans~A, we include the \Mmol\ from the CO sensitivity map for the resolved DGR. Also, including \Mmol\ mass limit (\alphacomw) has little to no effect on the DGR for this galaxy. In addition, the uncertainties in \Mdust\ and \MHI\ measurements are larger than the anticipated range of DGR variations derived from the CO sensitivity map. 

The grey density map in the background shows the literature trend with DustPedia+DGS galaxies from \citet{galliano2021nearby}. An anti-correlation between the two parameters has been shown, suggesting that when there are more stars per gas (inverse specific gas mass), it is easier for the gas to be enriched in dust (also in metal). The vertically extended DGR feature at a specific gas mass of $\sim$~0.1 is predominantly populated by early-type galaxies. 

As in the literature, an anti-correlation is also observed in the resolved specific gas mass and DGR. Another noticeable feature is that the more massive galaxies have shallower/flat trends than lower mass galaxies (NGC~1705 and Sextans~A). This is possibly consistent with the stochastic star-forming activities in dwarf galaxies. A smaller galaxy has a low average SFR, but the rate fluctuates widely between bursts and quiescent levels as cells individually and in small groups experience star formation.

\subsection{The resolved DGR\---\Zgas\ relation}
\label{sec:resolved_dgr_zgas}

\begin{figure*}
    \centering
    \includegraphics[scale=0.73]{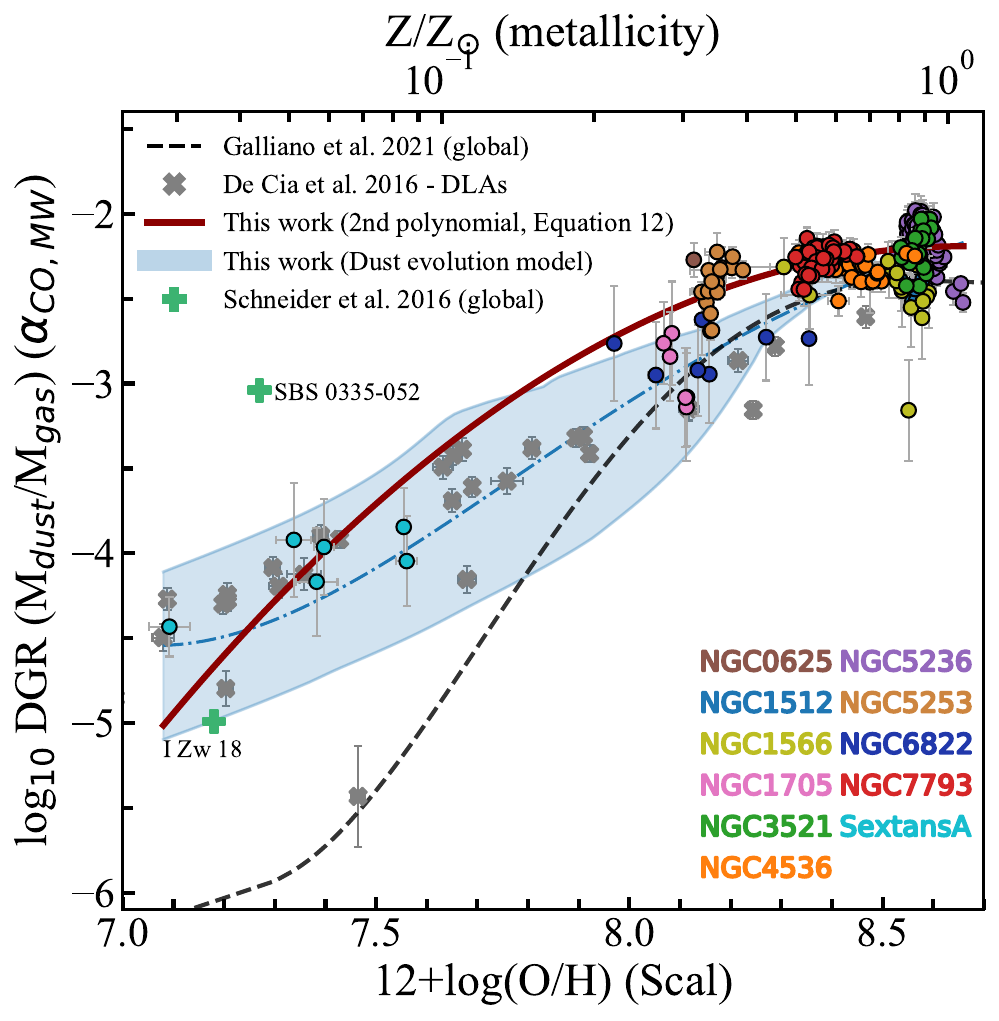}
    \caption{The resolved DGR\---\Zgas\ relationship of TYPHOON galaxies (colour-coded by galaxy) and the global relationship from references: The 4th polynomial fit for nearby DustPedia+DGS galaxies from \citet{galliano2021nearby} (black dashed line). The DGR from depletion methods with DLAs at higher-$z$ is from \citet{decia2016dust} (grey cross symbols). The red solid line presents the best fit for resolved TYPHOON galaxies. The blue-shaded area shows the dust evolution model (\texttt{BEDE}) runs with dust-related parameters at 16\--84th probability distribution. The blue dashed line presents the 4th polynomial fit for the models. Green plus symbols are the global DGR measurements for dwarf galaxies with different gas densities studied in \citet{schneider2016dust}. The colour-coded symbols are the same as previous figures. }
    \label{fig:discussion}
\end{figure*}

Figure~\ref{fig:discussion} presents the resolved DGR\---\Zgas\ relationship of TYPHOON galaxies, which generally follows the global trend (except for regions in Sextans~A). The scatter is large in both DGR and \Zgas\ as we can capture fluctuations at (sub-)kpc scale. One of the most notable features from our measurements is that the several star-forming regions in Sextans~A have DGR values of 0.5\--2~dex higher than the literature values at fixed metallicity. Additionally, while the resolved \Zgas\ values of this galaxy are spread across $\sim$~1~dex, implying chemical inhomogeneity within the galaxy, DGR values are relatively homogeneous given that the DGR varies only within $\sim$~0.5~dex. The wide range of \Zgas\ (7.1 $<$ \logOH\ $<$ 7.7) can be interpreted with a few distinguishable H~II regions having different star formation histories in Sextans~A investigated by \citet{gerasimov2022stellar}: northeast (NE) and southwest (SW) regions; see Figure~\ref{fig:SextansA} for metallicity spatial distribution. The two regions have different metallicity ranges such as 7.3 $<$ \logOH $<$ 8.1 and 7.0 $<$ \logOH $<$ 7.2 for NE and SW regions, respectively. Regarding star formation history, the NE region has had the most energetic star formation activities in the galaxy since $\sim$~200~Myr ago and already started to be enriched in metal abundance presenting a higher \logOH. In contrast, for the SW region, star-forming activity has likely started very recently ($\sim$~20~Myr ago), having a lower \logOH\ (\citealt{boyer2009spitzer}).

A large scatter stands out at the intermediate or critical metallicity regime (0.05\---0.3~$Z/Z_{\odot}$; 7.5 $<$ \logOH\ (Scal) $<$ 8.2), agreeing with the chemical evolution models (e.g., \citealt{asano2013dust} and \citealt{devis2017using}). In the theoretical view, the DGRs in this regime are predicted to rapidly increase as a function of metallicity compared to those in the high-/low- metallicity regimes. Specifically, in this critical metallicity regime, the dust evolution mechanism starts to be more efficient, as the dust growths in the ISM begin to contribute. The critical metallicity of a system can vary and be determined by the system's characteristics, such as star-formation timescale at a given time (t), $\tau_{\rm SF}$ = $M_{\rm ISM}$ (t) /SFR (t). If the $\tau_{\rm SF}$ is short, rapid star formation processes lead to early-stage metallicity increases, thereby accelerating the enrichment of dust with metals compared to scenarios with longer $\tau_{\rm SF}$ (\citealt{asano2013dust}; \citealt{feldmann2015equilibrium}).

We provide a best-fit model that describes the observed resolved DGR\---\Zgas\ relation by applying the n-th polynomial models. We increase `n' by up to 5 and to find what degree of polynomial fit best describes the observed DGR\---\Zgas\ relation, we employ Bayesian Information Criteria (BIC). With an assumption of Gaussian posterior probability, the BIC can be described below:
\begin{equation}
    BIC = \chi^2 + k \ln(n),
\end{equation}
where $\chi^2$ is the sum of squares of residuals from maximum likelihood, $\rm k$ is the number of model parameters, and n is the sample size (in our case, n = 346). In the process of calculating $\chi^2$, we use the Orthogonal Distance Regression (ODR) to find the best-fit model at a given degree of polynomial fit, taking the uncertainties for both the x-axis (\logOH) and y-axis (DGR) into account. The derived best-fit model is shown as a dark red solid line in Figure~\ref{fig:discussion}. The derived best fit for this study (Scal, \alphacomw) is expressed as below:
\begin{equation}
    \log_{10} {\rm DGR} = -1.1381~Z_{\rm gas}^2 + 19.6996~Z_{\rm gas} -85.4377.
\end{equation}

\section{Discussion}
\label{sec:discussion}

\subsection{Understanding the DGR\---\Zgas\ relationship}
\label{sec:discussion_depletion}

\subsubsection{Emission vs. absorption line methods}
Recently, many studies have reported tension on the global DGR\---\Zgas\ relationship, contrasting emission line-based methods (utilizing far-IR, H{\sc i}, CO, and optical emission line observations) with absorption line-based methods (employing metal depletion measurements from DLAs) (e.g., \citealt{decia2016dust}; \citealt{roman-duval2022metal3}). Emission line studies often show a change in slope at a certain metallicity, indicating a different mechanism of dust formation. In contrast, absorption line studies show a constant increase of DGR with \Zgas\ through the large range of \Zgas. Figure~\ref{fig:discussion} shows the DGR\---\Zgas\ relation derived from both methods: 1) emission line observations (global DGR\---\Zgas\ relation; \citealt{galliano2021nearby}) and 2) absorption line-based depletion method (\citealt{decia2016dust}).

Numerous factors contribute to this tension such as uncertainties both in emission and absorption line-based measurements as discussed in \citet{roman-duval2022metal3} and \citet{roman-duval2022metal4}. Emission line-based studies possess uncertainties such as biases in \Mdust\ measurements due to different assumptions on the far-IR opacity of dust in SED modelling. Plus, \Mdust\ can be underestimated when it is measured in galaxy-integrated scale (the `Matryoshka' effect that is introduced in \citet{galliano2018interstellar}; also see Section~\ref{sec:introduction}). Separating cold regions from warmer regions can be achieved at the finer resolution, resulting in systematically higher \Mdust\ (by up to 50\% higher than galaxy-integrated photometry method) as cold/large dust grains occupy the most \Mdust\ obtained. This has been observed in several studies, for example, in \citet{galliano2011non}, \citet{galametz2012mapping}, \citet{roman-duval2014dust}, and \citet{aniano2020modeling}. The systematic offset disappears when probed at a few tens parsec scale, which means this high resolution is required for better capturing the cold \Mdust. Our spatially resolved measurements, while still limited by resolution (SPIRE~350~$\mu$m or H{\sc i}\--21~cm), mitigate this effect, providing relatively accurate \Mdust\ with a median physical resolution of 0.594~kpc ranging from 0.200~kpc (NGC~6822 and Sextans~A) to 2.279~kpc (NGC~1512). However, \texttt{MAGPHYS} (and most SED modelling codes) assume a fixed dust emissivity index (i.e., the slope of the RJ tail), and this can introduce systematic biases on the resulting \Mdust\ values in the absence of longer wavelength data (\citealt{galliano2018interstellar}).

The aperture size chosen is also an important factor in uncertainties of emission line-based methods when it is galaxy-integrated studies, given different distributions of ISM phases (\citealt{leroy2008star}; \citealt{munozmateos2009}) (not only the DGR\---\Zgas\ relation but also other ISM scaling relations). Typically, stars, dust, molecular gas, and metals are concentrated within the optical radius ($r_{25}$) in star-forming galaxies, while atomic gas is extended further, up to 2\--4 times the optical radius (\citealt{hunter1997star}; \citealt{wang2013bluedisks}; also see Figure~\ref{fig:NGC0625}\--\ref{fig:SextansA}). Therefore, characterizing a galaxy's atomic gas mass for global DGR\---\Zgas\ relationships and other ISM scaling relations often involves spatially resolved \MHI\ maps from interferometric radio observations and sum up the \MHI\ within the chosen aperture size (\citealt{devis2017using}; \citealt{devis2019}) or an indirect estimation of \MHI\ within the optical radius from integrated H{\sc i} spectra from single-dish radio telescopes (\citealt{casasola2020ism}; \citealt{li2023interstellar}\footnote{Using observational evidence of the similar shapes in the radial profile of H{\sc i} surface density normalised by H{\sc i} radius, $R_{\rm HI}$ and the H{\sc i} size-mass relation (e.g., \citealt{wang2014observational}; \citealt{wang2020late})}). This effect becomes more noticeable in dwarf irregular galaxies such as NGC~6822 and Sextans~A of our sample, where the H{\sc i} gas morphology differs from the distribution of metals, stars, and molecular gas components. However, this issue is less pronounced at the spatially resolved scale due to the criteria ensuring reliable mass measurements with $S/N\geq3$. 

There are also uncertainties in absorption line-based methods. Many DGR and metallicity measurements from the depletion pattern investigation using absorption lines are carried out outside the MW (e.g., Large/Small Magellanic Cloud), but are still based on the MW depletion pattern. However, this assumption may not always hold (see Section 3.3.2 in \citealt{galliano2018interstellar}). The depletion methods for the DGR measurements assume metal depletion onto dust and abundance ratios, which are highly uncertain. In particular, the depletion measurements of carbon and oxygen are challenging because their UV transitions are easily saturated or too weak. Due to the limited measurements of C and O depletion outside the MW, the depletion method to estimate the DGRs of the system relies on the MW relation between the depletion of Fe and C (or O) (\citealt{decia2016dust}; \citealt{peroux2020cosmic}; \citealt{roman-duval2022metal3}). Additionally, the dust depletion factor [Zn/Fe] is frequently used to measure dust abundance, given that Zn is a volatile and Fe is a refractory element. However, this assumption is still debated (see Section 4.1 in \citealt{roman-duval2022metal4} and references therein).

Given these uncertainties in play, some of which are addressed in this study, our resolved DGR\---\Zgas\ relationship seems aligned with findings from the depletion method for DLAs. Particularly in the low metallicity range, areas within Sextans A show DGR values similar to those determined by the depletion method. Nonetheless, the fact that molecular gas has not been factored into the total \Mgas\ for these regions and that only one galaxy has been considered can be problematic when concluding. In the subsequent section, we discuss more on the DGR values in the low metallicity range.

\subsubsection{Low metallicity regime - Sextans~A}
\label{sec:sextansA}
Detection of CO lines in radio has rarely been successful in low metallicity systems. They can be simply H{\sc i}-dominant systems but at the same time it could be because they tend to favour CO-dark or CO-free molecular gas (\citealt{bolatto2013}). In such systems, the more effective penetration of high-energy UV photons through the low-metal/dust environment can consequently lead to the photodissociation of CO molecules. CO observations thus fail to trace molecular gas components optimally in the low metallicity systems. Also for this reason, the metallicity dependant $\alpha_{\rm CO}$ (i.e., \alphacoz) has been suggested for the past several years but the uncertainty on it is still substantial (e.g., \citealt{madden2012low}; \citealt{bolatto2013} for a review).

The reliable molecular mass measurement is challenging in the case of Sextans~A among our sample. To handle this issue, many studies including \citet{remyruyer2014}, \citet{casasola2020ism} and \citet{galliano2021nearby} used the ISM scaling relations. The relation between global \Matom/\Mstar\ and \Mmol/\Matom\ suggested by \citet{casasola2020ism} has been used for low metallicity DGS galaxies in \citet{devis2019} and \citet{galliano2021nearby} to infer their global \Mmol\ within the optical radius of low-metallicity galaxies. We thus adopt this scaling relation but in a resolved scale (see Figure~\ref{fig:sr_casasola20}). Given the \Matom/\Mstar\ range of individual regions in Sextans~A, the \Mmol/\Matom\ is inferred and applied to the DGR of the corresponding regions. We see only a negligible change in DGRs for regions in Sextans A, far smaller than the error itself thus we do not show them in the figure. It should be noted that the \Mmol/\Matom\ we derived for other galaxies are derived from the \alphacomw, which might possess a possibility for \Mmol\ being underestimated.

In \citet{cigan2016herschel}, they used one of the far-IR fine spectral lines, [C~II]~158~$\mu$m to estimate the \Mmol\ in the ISM under an assumption that [C~II], as an important coolant of the ISM in such low-metallicity systems, can trace the H$_2$ in photodissociation regions (PDRs). More recently, \citet{cigan2021herschel} has inferred that the \Mmol/\Matom\ for Sextans~A is $\sim$~4, from the [C~II] line detected in this galaxy. We consider this molecular gas fraction and estimate the `conservative' gas masses and DGRs of all regions in Sextans~A. DGR values after applying \Mmol/\Matom\ of $\sim$~4 are naturally much lower (by 0.7~dex). However, the DGR values from [C~II] line may be overestimated because [C~II] lines trace molecular gas in PDRs, including diffuse atomic and diffuse molecular gas, which are more extended than star-forming dense molecular gas regions we probe in this relation. Considering the conservative \Mmol\ for regions in Sextans~A returns that the DGRs of a few regions (decreased by 0.7~dex) agree well with the literature trend from global trend (\citealt{galliano2021nearby}), however, the DGR values of most Sextans A regions still show being large (up to $\sim$~1~dex than the literature).

The high DGR values observed in Sextans A can be attributed to the intrinsic large scatter of DGR in low-metallicity regimes. One of the factors is the gas density, which has been reported in \citet{schneider2016dust} for two galaxies, SBS~0335-052 (\logOH\ $\sim$~7.27) and I~Zw~18 (\logOH\ $\sim$~7.18). SBS~0335-052 has a significantly higher global DGR up to 3~dex than I~Zw~18 even at similar metallicity (see green plus symbols in Figure~\ref{fig:discussion}). \citet{schneider2016dust} interpret this discrepancy as a result of differences in gas densities of the two galaxies, suggesting that the high molecular gas density in SBS~0335-052 ($n_{\rm mol}$~$\sim$~1500~cm$^{-3}$) enables grain growth process in the ISM, which is more efficient dust mass build-up compared to condensation in SNe, despite its low metallicity. On the other hand, I~Zw~18 has a very low gas density ($n_{\rm mol}$~$\sim$~91~cm$^{-3}$), 20 times lower than SBS~0335-052. \citet{schneider2016dust} hint that the factor that determines the DGR of a system is not only the metallicity but also the gas density. This agrees with \citet{chiang2018spatially} where the increasing trend of dust-to-metal mass ratio (DMR $\equiv$ \Mdust/$M_{\rm metals}$) as a function of molecular gas fraction has been shown in M~101. Furthermore, the argument is supported by the depletion method in recent work by \citet{hamanowicz2024metal}, where they found a trend of DGR values increasing with the higher hydrogen column density, N(H), in several absorption lines of a few sight-lines towards Sextans~A. They found that at high N(H) ($>$ 10$^{21}$~cm$^{-2}$), the DGR values are in line with the high DGR values from our work, whereas at N(H) $\sim$ 10$^{20}$~cm$^{-2}$, the DGRs are reduced by 1 dex, reaching the low value found by the literature we are referring to (\citealt{galliano2021nearby}).

Other factors are possible to explain the high DGR values in the low-metallicity regime, such as higher dust yield from SNe, a minor fraction of dust destruction by SNe, or being affected less by photofragmentation, which will be covered by model fitting in Section~\ref{sec:discussion_model}.

\subsubsection{Broken vs. single power law}

We briefly explore the optimal shape of the observed DGR\---\Zgas\ relation, contrasting single- (\citealt{devis2019}) or broken- power laws (\citealt{remyruyer2014}). It should be highlighted that different authors use different metallicity calibrations so a fair comparison is challenging. With this caveat in mind, we compare the broken-power law relation by \citet{remyruyer2014}, where PT05 metallicity calibration is used (see Section~\ref{sec:introduction}) and the single-power law relation by \citet{devis2019}, where Scal is used. To more fairly compare both relations, we also calculate the resolved DGR\---\Zgas\ relation with PT05 metallicity calibration for our galaxies (see the rightmost panel in Figure~\ref{fig:different_diagnostics}).

\begin{figure}
    \centering
    \includegraphics[scale=0.48]{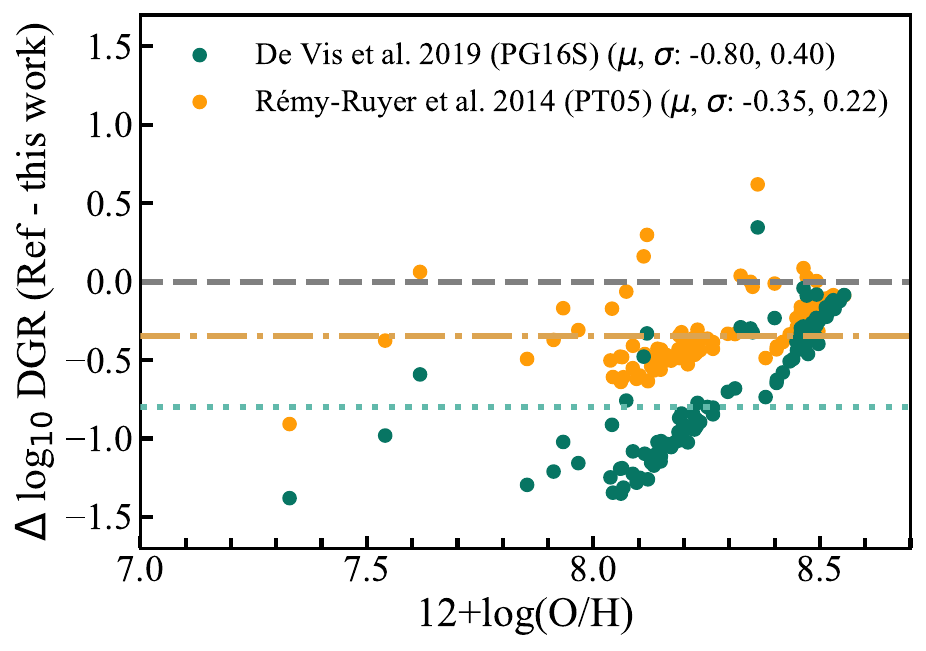}
    \caption{The residual of DGR values between the two reference models using the consistent metallicity calibration: the residuals with the single power law from \citet{devis2019} (Scal; green) and those with the broken power law from \citet{remyruyer2014} (PT05; orange). Each median difference of DGRs is indicated as dotted and dash-dotted lines, respectively.}
    \label{fig:discussion_devDGR}
\end{figure}

In Figure~\ref{fig:discussion_devDGR}, we show the deviation between the literature DGRs and the observed DGRs at a fixed \Zgas, including the median difference and its standard deviation. Our resolved DGRs are systematically higher than the two galaxy-integrated studies, hinting that the effect on the chosen aperture size when deriving the ISM masses is mitigated in our study. The broken power law proposed by \citet{remyruyer2014} has a smaller residual and standard deviation than the single power law by \citet{devis2019}.

This finding aligns with similar studies by \citet{vilchez2018metals} on M~101 and NGC~628 at a spatially resolved scale, where the broken power-law has been suggested for describing the DGR\---\Zgas\ relation. However, it is difficult to make a definitive conclusion about which functional form better explains the measurements due to the limited number of low-metallicity regions and the lack of \Mmol\ information in those systems. When considering dust evolution models proposed in previous studies (e.g., \citealt{asano2013dust}; \citealt{feldmann2015equilibrium}), and to be discussed in Section~\ref{sec:discussion_model}, there is a preference for higher-order relationships over linear functions. On the other hand, given the smaller residuals of the broken power law compared to the single power law in intermediate-to-high metallicity regimes (\logOH $>$ $\sim$8.0), the inclusion of molecular gas information for Sextans~A would still support deviations from the single linear relation.

\subsection{Dust evolution models}
\label{sec:discussion_model}
We implement a grid-based chemical evolution model to analyse the observed resolved DGR\---\Zgas\ relationship. We utilize a Bayesian Monte Carlo Markov Chain (MCMC) based one-zone dust evolution model, \texttt{BEDE}\footnote{\url{https://github.com/zemogle/chemevol}}, developed by \citet{devis2021bede} (also see \citealt{rowlands2014dust} and \citet{devis2017using} for the initial approach). We fit the model to the entire resolved data points from all galaxies simultaneously.

\begingroup
\setlength{\tabcolsep}{4pt}
\begin{table}
    \centering
    \begin{tabular}{c|c|c}
        \hline
        \hline
        Parameter & Description & Prior range\\
        \hline
        $SN_{\rm red}$ & Factor dust yields for SNe are reduced by & 1\---5 \\
        $k_{\rm frag}$ & Photofragmentation of dust grains & 0.003\---0.5 \\
        $k_{\rm gg,cloud}$ & Grain growth in the clouds & 1000\---16000 \\
        $f_{\rm dif}$ & Maximum dust-to-metal ratio in diffuse ISM & 0.2\---0.4 \\
        \hline
        $M_{\rm destr}$ & \Mdust\ destructed per SN & 15 \\
        $k_{\rm gg,dif}$ & Grain growth in the diffuse ISM & 5 \\
        \hline
        \hline
    \end{tabular}
    \caption{Dust-related parameters and their prior range used in this study. We fixed the $M_{\rm destr}$ and $k_{\rm gg,dif}$ to the reference values from \citet{devis2021bede} as they are found not to significantly change DGR values.}
    \label{tab:bede}
\end{table}
\endgroup

\begin{figure}
    \centering
    \includegraphics[scale=0.55]{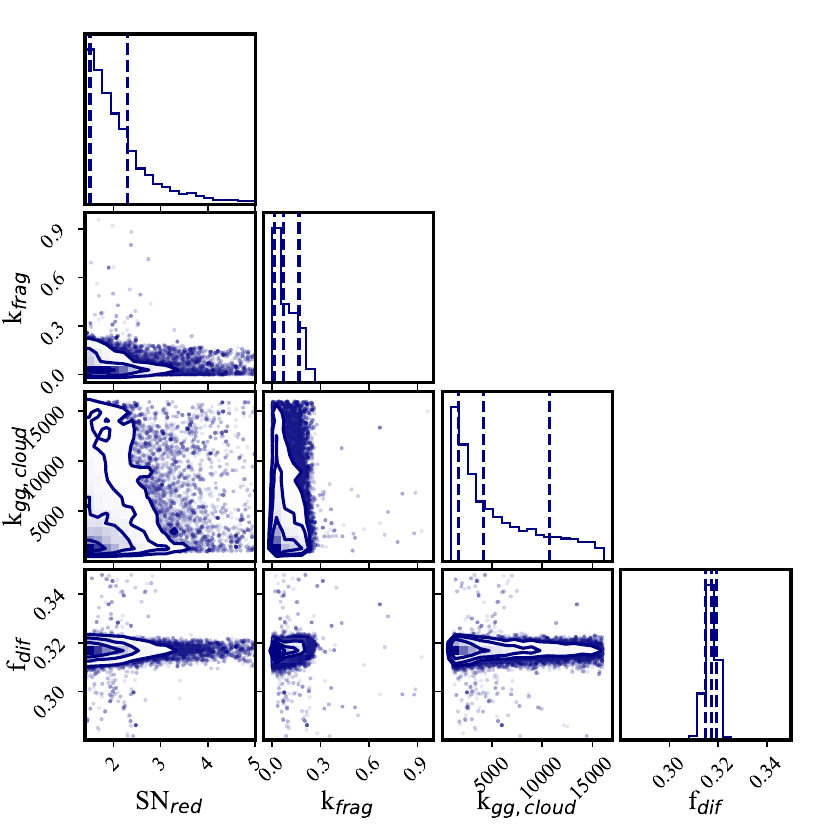}
    \caption{Probability distribution for the MCMC run for dust-related parameters. Each dashed line from the left shows the distribution's 16th, 50th and 84th percentile. The probability distribution is well constrained generally, but $k_{\rm frag}$ shows a feature reached to the lower bound (see Section~\ref{sec:model_interpret}). The values for each percentile are presented in Table~\ref{tab:bede_result}.}
    \label{fig:corner}
\end{figure}

\subsubsection{Model setting}
\label{sec:model_setting}
\texttt{BEDE} integrates various factors governing the evolution of gas, stars, metals, dust, and SFR over time, assuming user-defined basic parameters (e.g., total mass, stellar IMF, star formation history, metal yields, inflow/outflow rate). We create a grid of total mass ($M_{\rm tot}$) with a range of $8 < {\rm log} M_{\rm tot}$/\Msol\ $< 10$ with a step size of 0.5 in log space, and fix the following: 1) stellar IMF to \citet{chabrier2003} IMF for consistency with the SED fitting approach (see Section~\ref{sec:magphys}), 2) star formation history to have a continuous increase of star formation efficiency (SFE) at an intermediate pace (SFE$_0$ = 10$^{-9}$~yr$^{-1}$ in Equation 9 of \citealt{devis2021bede}). 3) The metal yield tables for SNe and AGB stars are set to \citet{limongi2018presupernova} with a stellar rotational velocity of 150~\kms, and \citet{karakas2018heavy} with high mass loss rates. The inflow rate is set to half of the $M_{\rm tot}$ which means the system has half of the $M_{\rm tot}$ at the beginning of its evolution and the other half amount is accreted from outside. The age grid of each modelled system starts from 0 to 13.8~Gyr with steps of 0.3~Gyr. \texttt{BEDE} is based on the THEMIS dust model (\citealt{jones2013evolution}; \citealt{kohler2015dust}; \citealt{jones2017global}), taking the dust grains made up of the mixture of carbon and silicate and allowing the grain size variation. For consistency with the observed dust mass from \texttt{magphys}, we increase the dust mass from models by 1.075 (\citealt{devis2019}).

We mainly focus on parameters whose variation affects the shape of the DGR\---\Zgas\ relation noticeably. The dust abundance in the model run is regulated by six parameters, which we term `dust-related parameters', including $SN_{\rm red}$, $k_{\rm gg, cloud}$, $k_{\rm frag}$, $f_{\rm dif}$, $M_{\rm destr}$ and $k_{\rm gg,dif}$. In this study, we only tune the first four parameters as the others marginally change the DGR\---\Zgas\ relation (see Figure 6 in \citealt{devis2021bede}), and at the same time disregarding two parameters reduces the computational time significantly. Table~\ref{tab:bede} presents the dust-related parameters and their prior range. We briefly present the general form of dust mass evolution in Appendix~\ref{app:model_description}. Additionally, we use the statistical frameworks outlined in \citet{devis2021bede} for a Bayesian MCMC approach. For the full description of the models, we refer readers to \citet{devis2021bede}.

The model DGR value of a system from the model runs is calculated by ($M_{\rm dust}$/$M_{\rm gas}$; Equation~\ref{eq:dgr}), and \Zgas, or \logOH\ is measured by the equation following \citet{devis2017using} and \citet{devis2021bede} as below:
\begin{equation}
    12 + {\rm log(O/H)} = 12 + {\rm log}\left(\frac{(M_{\rm O}-0.238 \times M_{\rm dust})/16}{M_{\rm gas}/1.36} \right),
\end{equation}
where $M_{\rm O}$ is the total oxygen mass in the system, and 0.238 is to subtract the oxygen content in the dust component. 16 is the atomic weight of Oxygen, and 1.36 is to subtract the helium abundance of the gas. The initial value of each parameter is chosen uniformly from the prior range. We use 50 walkers and let them move 1000 steps. We disregard the first 50 steps as burn-in walks.\footnote{To ensure the step size is enough for the walkers to cover the entire prior range, we estimate the integrated autocorrelation time and acceptance fraction using the \texttt{emcee} packages. They are estimated as 14, and 0.445, respectively, which means the model run provides a good fit.} Figure~\ref{fig:corner} shows the probability maps of each combination of parameters after the MCMC run. The median (50\%), 84th percentile, and 16th percentile are presented in Table~\ref{tab:bede_result}.

\begingroup
\setlength{\tabcolsep}{10pt}
\renewcommand{\arraystretch}{1.5}
\begin{table}
    \centering
    \begin{tabular}{c|c}
        \hline
        \hline
        Parameter & Probability (50th$^{\rm 84th}_{\rm 16th}$) \\
        \hline
        $SN_{\rm red}$ & 1.50$^{+0.80}_{-0.40}$ \\
        $k_{\rm frag}$ & 0.07$^{+0.09}_{-0.06}$ \\
        $k_{\rm gg,cloud}$ & 4131.81$^{+6562.80}_{-2550.28}$ \\
        $f_{\rm dif}$ & 0.32$^{+0.00}_{-0.00}$ \\
        \hline
        \hline
    \end{tabular}
    \caption{The MCMC run result for dust-related parameters with the 16, 50, and 84th percentile of the probability.}
    \label{tab:bede_result}
\end{table}
\endgroup

\subsubsection{Interpretations and comparison with other studies}
\label{sec:model_interpret}
In Figure~\ref{fig:discussion}, we show the \texttt{BEDE} results with a blue shaded region, run with 16\%\--84\%\ of each parameter. The non-dust-related parameters are set the same as in Section~\ref{sec:model_setting}. We provide the 4-th polynomial fit to the best-fit models on the resolved DGR\---\Zgas\ relation as below, for a convenient comparison with \citet{galliano2021nearby} where the same order of polynomial fit is suggested for the galaxy-integrated properties of local galaxies.

\begin{align}
    \begin{split}    
        &\rm \log_{10} {\rm DGR} = 0.4426~Z_{\rm gas}^4 - 14.7297~Z_{\rm gas}^3 + 183.02~Z_{\rm gas}^2 \\
        &\rm ~~~~~~~~~~~~~~~~~~~~~ - 1004.77~Z_{\rm gas} + 2050.49
    \end{split}
    \label{eq:bestfit_model}
\end{align}

In \texttt{BEDE}, the relationship between DGR and metallicity is characterized by a metallicity threshold at around \logOH\ of 7.7\--8.2. Above this threshold, dust formation and destruction processes reach equilibrium so that a linear relation between DGR and metallicity is expected.
\begin{enumerate}
    \item[-] {Out of the four dust-related parameters,} $f_{\rm dif}$ is related to the DMR of the system, indicating the available amount of metals to build up dust (DMR; $\neq$ $f_{\rm dif}$) \footnote{DMR definition varies in different references. Our paper defines DMR as \Mdust/$M_{\rm metals}$. The $f_{\rm dif}$ corresponds to \Mdust/($M_{\rm metals} + $\Mdust).}, and it determines the DGR range above the metallicity threshold. In our model run, we found that the model runs well converged to $f_{\rm dif} = 0.32$. This value is greater than the $f_{\rm dif}$ found in \citet{devis2021bede}, where dust evolution modelling is carried out on 340 late-type galaxies from the DustPedia, HIGH, and HAPLESS surveys \citet{devis2021bede}. They found $f_{\rm dif}$ of 0.204, agreeing with the observations ($= 0.214$) from the DustPedia galaxies in \citet{devis2019}. Our run of the dust evolution model suggests a maximum DMR of 0.47, which agrees with the theoretical maximum metal amounts locked in dust investigated by \citet{palla2024metal} (see Appendix~A within). The $f_{\rm dif}$ hardly affects DGR values below the metallicity threshold, as at lower metallicity or early stage of galaxy evolution, a negligible amount of metals in the ISM contributes to dust mass.
\end{enumerate}

Below the metallicity threshold in the models (i.e., \logOH\ $\leq$ 7.7\--8.2), the other three dust-related parameters, $SN_{\rm red}$, $k_{\rm gg,cloud}$, and $k_{\rm frag}$ affect the dust abundance as below:
\begin{enumerate}
    \item[-] $k_{\rm gg,cloud}$ is the scaling factor that reduces the timescale of dust grain growth in the dense ISM, implying that the higher $k_{\rm gg,cloud}$ the faster the system reaches the metallicity threshold (\citealt{asano2013dust}; \citealt{devis2017using}). The best fit for $k_{\rm gg,cloud}$ is 4131.81, consistent with other studies with DustPedia galaxies in galaxy-integrated scale. For instance, \citet{devis2021bede} found 3820 as the median with large uncertainties (16/84th percentile of $>$ 1500), and \citet{galliano2021nearby} showed that median $k_{\rm gg,cloud}$ of 4485 (16/84th percentile of $\sim 20$) describes the observed quantities of galaxies the best. The dust grain growth timescale spans 940~Myr in the low-metallicity system (at \logOH\ $\sim$ 7.08) and 10~Myr at \logOH $\sim$ 8.66, indicating a faster grain growth at near solar metallicity.
    \item[-] $k_{\rm frag}$ regulates the photofragmentation of large grains by UV radiation, impacting dust destruction in the diffuse ISM in the system. Higher values of $k_{\rm frag}$ decrease the photofragmentation timescale of large a-C:H/a-C\footnote{Amorphous carbon with partial hydrogenation} grains. Silicate grains are excluded from the models due to their resistance to photofragmentation.The median $k_{\rm frag}$ (0.07) is found low from the parameter space, which agrees with the model run for the galaxy-integrated scale (0.03 in \citealt{devis2021bede}). This low $k_{\rm frag}$ implies that photo-fragmentation marginally affects the dust grain destruction in the observed regions. It is also observed that the distribution seems to have reached a lower bound at 0. It will be outside the allowable range if it falls below this value.
    \item[-] $SN_{\rm red}$ is another dust-related parameter that scales down the dust yield for SNe (see \citealt{todini2001dust} or Table 5 in \citealt{galliano2021nearby}). This parameter has a substantial impact on DGR values, particularly in low-metallicity systems where stardust is the primary dust production mechanism; higher $SN_{\rm red}$ leads to lower DGR values. The lower bound of this parameter, where $SN_{\rm red} = $ 1 means 100\% of the produced dust from the SNe is preserved and contributes to the dust abundance in the system. Our model runs find $SN_{\rm red} = 1.50^{+0.80}_{-0.40}$, corresponding to 43\--91\% of SN dust yield (or, 0.43\--0.91 \Msol/SN when the star mass is 25~\Msol following \citealt{todini2001dust} dust yield table) is survived to build up the dust mass in the low-metallicity regions. This contrasts with \citet{galliano2021nearby}, where a very low dust yield from SNe ($\sim$~0.025 \Msol/SN) is found. However, our run has generally agreed with other studies such as in \citet{devis2021bede} ($SN_{\rm red}$ of 2.01), or \citet{nanni2020gas} ($>$ 25\% of dust condensation fraction from SNe). Also, \citet{delooze2020jingle} has retrieved high SN dust yields, similar to \citet{nanni2020gas}.
\end{enumerate}

\subsubsection{Caveats on our dust evolution modelling}
This model comparison has caveats such as the dust evolution model we utilized assumes one-zone chemical and dust evolution. Thus, it may not be suitable for spatially resolved studies as the chemical/dust evolution can differ depending on the regions as they have different environments in terms of spiral/bar structure, or inflow/outflow regions. 

Several degeneracies have been reported in \citet{devis2021bede}, such as the assumed IMF, recycling fraction, and SN yield tables. Also, it is worthwhile to note that the DGR values in the low-metallicity systems are regularised by $k_{\rm frag}$ and $SN_{\rm red}$ at the same time, which can lead to another degeneracy.

The combination of low $k_{\rm frag}$ and $SN_{\rm red}$ can reproduce the DGR range of Sextans~A. However, we found the $k_{\rm frag}$ reaches the lower bound, implying other parameters should also be considered. For example, in this model, the grain growth timescale in the dense clouds in such a low-metallicity regime is too long to contribute to dust abundance spontaneously. In reality, it could not always be the case given that several studies suggest that high gas density can trigger grain growth in the ISM in the early stage of galaxy evolution. However, we are cautious given that only one galaxy (Sextans~A) in the low-metallicity system is probed in this study, the best-fit parameter range found could be biased toward this galaxy.

\section{Summary}
\label{sec:summary}
In this study, we probed the DGR\---\Zgas\ relationship with 11 nearby galaxies from the TYPHOON survey in a spatially resolved manner at (sub-)kpc scale. With the large FoV of TYPHOON data, we map the gas-phase metallicity for $\gtrsim$~1 optical size of galaxies. We derived the DGR from the spatial resolution-matched multi-wavelength datasets (from far-UV to far-IR, CO and H{\sc i} radio). With these datasets, we investigate the resolved DGR\---\Zgas\ relationship. We summarise the result as follows:
\begin{enumerate}
    \item[-] The global DGR\---\Zgas\ relationship of TYPHOON galaxies follows well with the literature (e.g., \citealt{devis2019}; \citealt{galliano2021nearby}), except for the lowest metallicity galaxy, Sextans~A.
    \item[-] The resolved DGR\---\Zgas\ relationship is explored, focusing on Scal metallicity calibration and \alphacomw. The resolved relationship shows a scattered distribution around each galaxy's global measurement. The observed trend prefers the broken power law, agreeing with the previously studied resolved DGR\---\Zgas\ relation in other nearby galaxies, for example, M~101 and NGC~628 (\citealt{vilchez2018metals}).
    \item[-] The high DGR values (up to $\sim$~2~dex than the literature) for the resolved regions in Sextans~A are observed, and the large span of \Zgas\ (up to 1~dex). These values are comparable with the DGRs from the depletion method for DLAs.
    \item[-] We run the dust evolution modelling with the Bayesian MCMC approach, \texttt{BEDE} (\citealt{devis2021bede}), varying the SN dust yield, photofragmentation factor of dust grains, grain growth in the dense clouds, and maximum dust-to-metal ratio in the diffuse ISM. We then find the best parameter sets that describe the observed resolved DGR\---\Zgas\ relationship of TYPHOON galaxies.
    \item[-] From the model fitting, we showed that high SN dust yield (43\--91\%) is preserved and an ineffective photofragmentation of dust grains can explain the observed high DGR values of Sextans~A. We note that the two parameters controlling the two effects (SN dust yield and photofragmentation) have reached the bound of each prior range, indicating that other dust production parameters can be considered. A previous study on the global DGR\---\Zgas\ relation for two low-metallicity dwarf galaxies (\citealt{schneider2016dust}) has shown that the two galaxies with comparable metallicity but different gas densities have large differences in DGR. Given this, it is possible that the DGR values can be regulated not only by the metallicity but also by the gas density of the systems.
    \item[-] We found a large scatter for the parameter that controls the grain growth timescale in dense clouds, which is consistent with the global studies (e.g., \citealt{devis2021bede}). However, we also attribute it to the lack of sample points in the metallicity range where the parameter causes variance of DGR. Additionally, a high maximum dust-to-metal ratio in diffuse ISM is found (DMR $<$ 0.47), which agrees with the theoretical view.
\end{enumerate}

A major limitation of this study is that only one very low metallicity galaxy is available with the required data. An avenue to improve this is to use the Hi-KIDS survey\footnote{\url{https://hikids.datacentral.org.au/}}, which has conducted observations of over 80 dwarf galaxies using the KOALA IFU instrument at the Anglo-Australian Telescope (AAT), where all of these galaxies have ancillary H{\sc i} data and 12 of them have {\it Herschel} far-IR observations with sufficient S/N for characterising both (atomic) gas and dust mass. Furthermore, an extension program with AAT/KOALA (PID: A/2024A/13; PI: Park) has observed 4 more dwarf galaxies with H{\sc i} and {\it Herschel} far-IR data. We plan to probe the resolved DGR\---\Zgas\ in these galaxies, for a more comprehensive study on the low-metallicity regime in a future study.

\section*{Acknowledgements}
The authors thank the anonymous referee who has provided constructive and helpful comments to improve our paper.
H-JP thanks Bärbel Koribalski for providing LVHIS ATCA H{\sc i} data cubes for NGC~625, NGC~1512, NGC~1705, and NGC~5253. H-JP thanks Kelsey Johnson and Molly Finn for providing the NGC~7793 ALMA data cube and the moment~0 map. H-JP thanks WALLABY teams (P.I. Tobias Westmeier) for providing the NGC~1566 H{\sc i} data from the Early Science data products. 

LC acknowledges support from the Australian Research Council Discoverty Project funding scheme (DP210100337). KG is supported by the Australian Research Council through the Discovery Early Career Researcher Award (DECRA) Fellowship (project number DE220100766) funded by the Australian Government. BFM thanks The Observatories of the Carnegie Institution for Science for believing in and supporting this decades-long programme at the duPont telescope at Las Campanas, Chile. 

Parts of this research were supported by the Australian Research Council Centre of Excellence for All Sky Astrophysics in 3 Dimensions (ASTRO 3D), through project number CE170100013. 

Most of the photometric UV, optical, and IR data for individual galaxies are retrieved from the NASA/IPAC Infrared Science Archive (IRSA) and the NASA/IPAC Extragalactic Database (NED), both of which are funded by the National Aeronautics and Space Administration and operated by the California Institute of Technology. The Australia Telescope Compact Array is part of the Australia Telescope National Facility (\url{https://ror.org/05qajvd42}) which is funded by the Australian Government for operation as a National Facility managed by CSIRO. This scientific work uses data obtained from Inyarrimanha Ilgari Bundara / the Murchison Radio-astronomy Observatory. We acknowledge the Wajarri Yamaji People as the Traditional Owners and native title holders of the Observatory site. CSIRO’s ASKAP radio telescope is part of the Australia Telescope National Facility (\url{https://ror.org/05qajvd42}). Operation of ASKAP is funded by the Australian Government with support from the National Collaborative Research Infrastructure Strategy. ASKAP uses the resources of the Pawsey Supercomputing Research Centre. The establishment of ASKAP, Inyarrimanha Ilgari Bundara, the CSIRO Murchison Radio-astronomy Observatory and the Pawsey Supercomputing Research Centre are initiatives of the Australian Government, with support from the Government of Western Australia and the Science and Industry Endowment Fund. This study makes use of the following ALMA data:  ADS/JAO.ALMA\#2013.1.01161.S, ADS/JAO.ALMA\#2015.1.00121.S, ADS/JAO.ALMA\#2015.1.00782.S,  ADS/JAO.ALMA\#2015.1.00925.S, ADS/JAO.ALMA\#2015.1.00956.S, ADS/JAO.ALMA\#2016.1.00386.S, ADS/JAO.ALMA\#2017.1.00392.S, ADS/JAO.ALMA\#2017.1.00886.L, ADS/JAO.ALMA\#2018.1.00219.S,  ADS/JAO.ALMA\#2018.1.01651.S, ADS/JAO.ALMA\#2018.1.01783.S, ADS/JAO.ALMA\#2019.2.00110.S, ADS/JAO.ALMA\#2021.1.00330.S, ALMA is a partnership of ESO (representing its member states), NSF (USA) and NINS (Japan), together with NRC (Canada), NSTC and ASIAA (Taiwan), and KASI (Republic of Korea), in cooperation with the Republic of Chile. The Joint ALMA Observatory is operated by ESO, AUI/NRAO and NAOJ. The National Radio Astronomy Observatory is a facility of the National Science Foundation operated under a cooperative agreement by Associated Universities, Inc. This work has made use of data from the European Space Agency (ESA) mission {\it Gaia} (\url{https://www.cosmos.esa.int/gaia}), processed by the {\it Gaia} Data Processing and Analysis Consortium (DPAC, \url{https://www.cosmos.esa.int/web/gaia/dpac/consortium}). Funding for the DPAC has been provided by national institutions, in particular the institutions participating in the {\it Gaia} Multilateral Agreement.

This work made use of Astropy:\footnote{http://www.astropy.org} a community-developed core Python package and an ecosystem of tools and resources for astronomy \citep{astropy2013, astropy2018, astropy2022}. This work also used Numpy (\citealt{numpy}), Matplotlib (\citealt{matplotlib}), and Scipy (\citealt{scipy}).

This research was conducted on Ngunnawal Indigenous land.

\section*{Data Availability}

The TYPHOON optical IFS datacubes will be made publicly available in a forthcoming release by Seibert et al. (in prep.) through Data Central\footnote{\url{https://datacentral.org.au/}}. The emission line data products can be made available upon reasonable request by emailing A. Battisti.


\bibliographystyle{mnras}
\bibliography{references} 



\appendix

\section{The DGR\---\Zgas\ relationship with different metallicity calibrations}
\label{app:diff_calibrations}

\begin{figure*}
    \centering
    \includegraphics[scale=0.35]{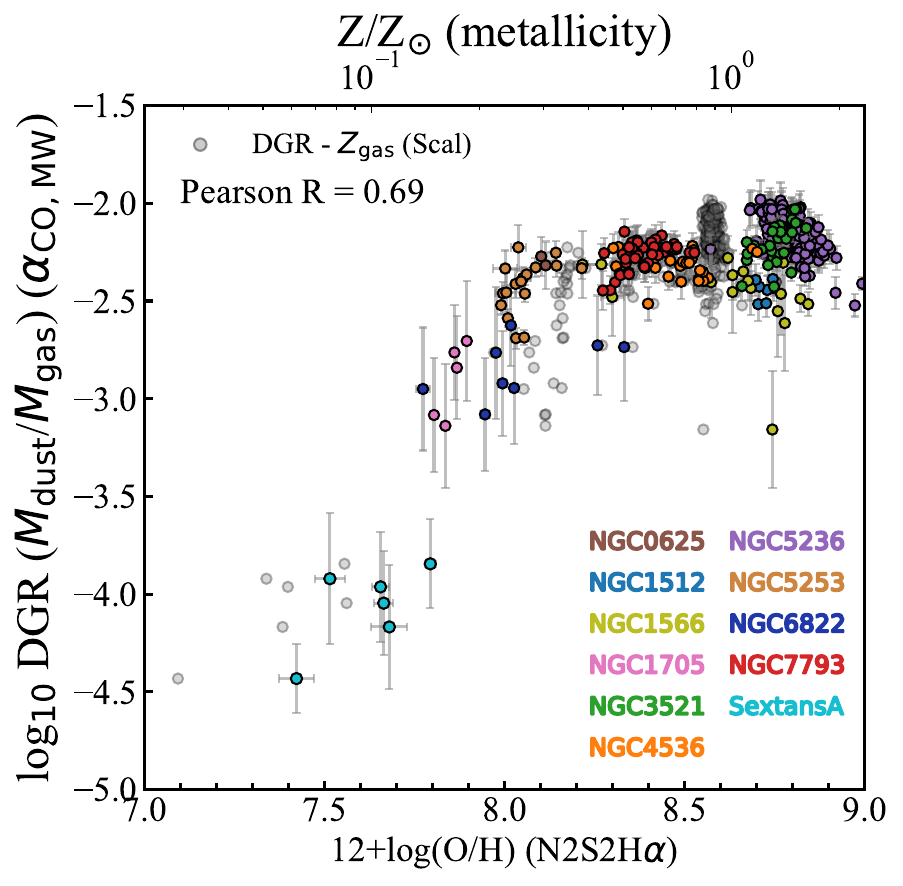}
    \includegraphics[scale=0.35]{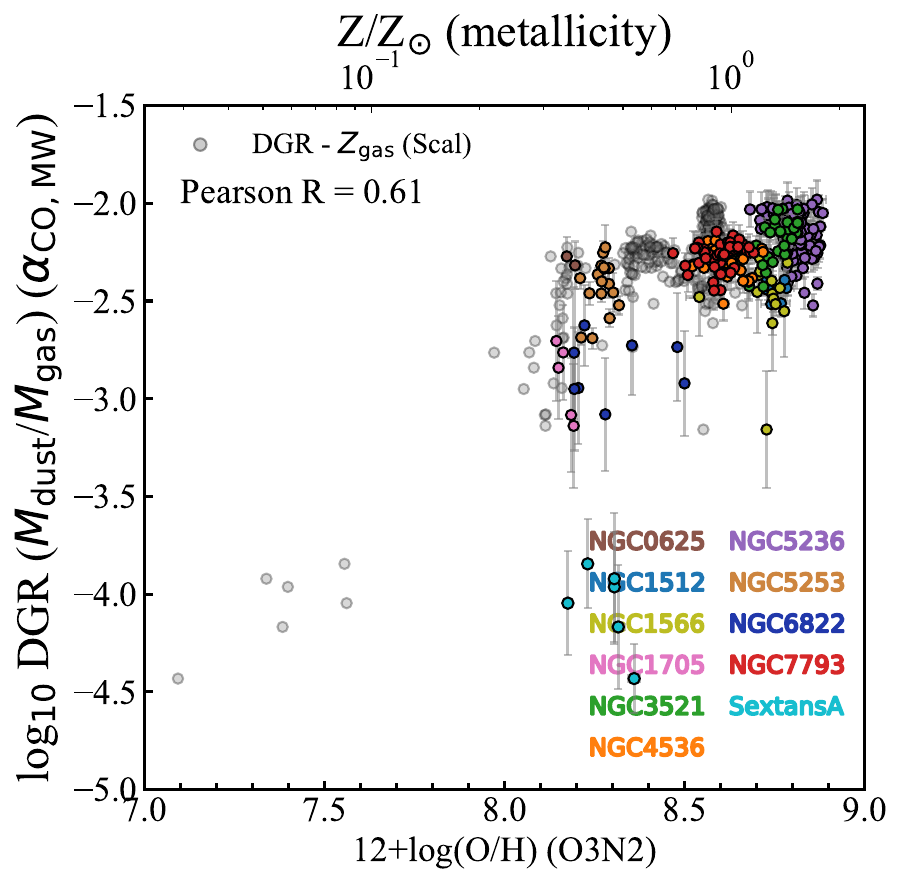}
    \includegraphics[scale=0.35]{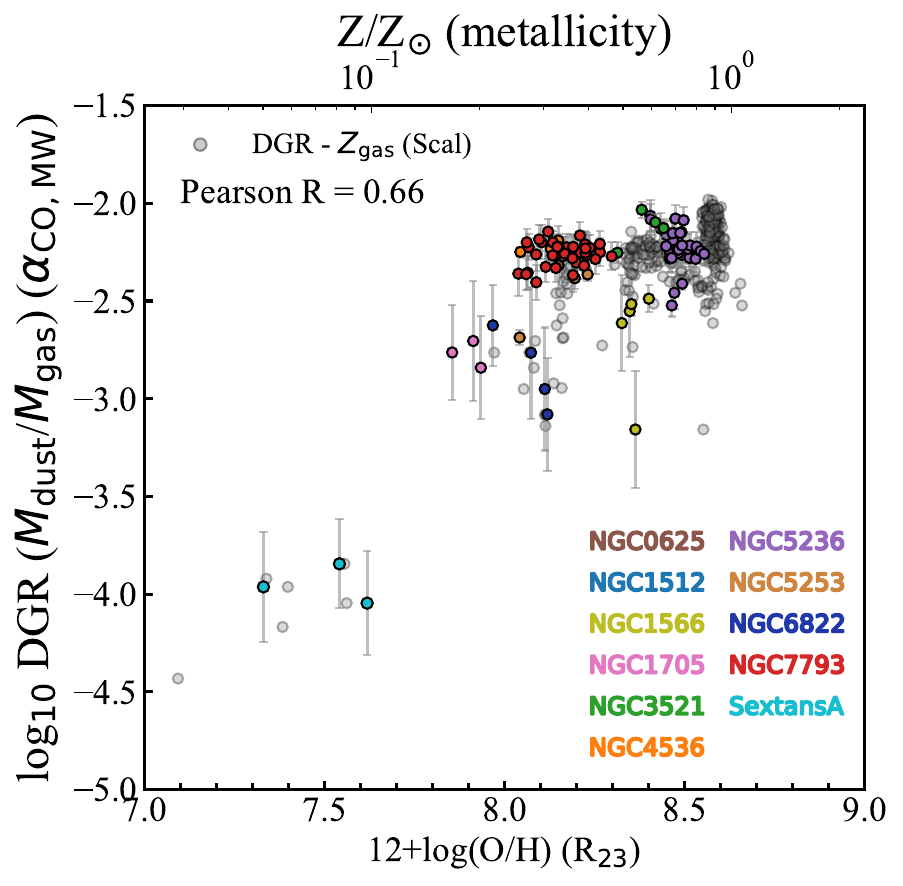}
    \caption{The resolved DGR\---\Zgas\ relationship with different metallicity diagnostics (N2S2H$\rm \alpha$; \citealt{dopita2016}, O3N2; \citealt{pettini2004abundance}; R$_{\rm 23}$ or PT05; \citealt{pilyugin2005oxygen}). The molecular gas mass has been measured in an assumption of \alphacomw\ for the resolved DGR values. The background grey dots present the relationship with Scal adopted in this study for reference.}
    \label{fig:different_diagnostics}
\end{figure*}

We show the resolved DGR\---\Zgas\ relationship using different gas-phase metallicity diagnostics, N2S2H$\alpha$, O3N2, and R$_{23}$ (Figure~\ref{fig:different_diagnostics}). We argue that the user should be aware that each calibration shows a different DGR\---\Zgas\ relation, especially at a low-metallicity system. Note that the DGR measurements include molecular gas estimation using \alphacomw.

{\it N2S2H$\alpha$}: The N2S2$\alpha$ calibration (\citealt{dopita2016}) uses combination of emission lines in the form of [N~II]$\lambda$6583/[S~II]$\lambda$$\lambda$6717,6731 and [N~II]$\lambda$6583/H$\alpha$. This calibration is one of the most popular diagnostics since the lines are close enough, allowing us to ignore the dust extinction effect. Additionally, the N2S2H$\alpha$ method is insensitive to the ionization parameter and the ISM pressure (P) within the range of 4 $<$ log(P/k) $<$ 7. Moreover, \citet{yates2021lgalaxies} presented that the gas-phase metallicity for low-mass galaxies obtained by the N2S2H$\alpha$ calibration is in good agreement with the direct $T_{\rm e}$ method. At the same time, the diagnostic derives super-solar gas-phase metallicity for very massive sources.

{\it O3N2}: The gas-phase metallicity from O3N2 (\citealt{pettini2004abundance}) has a very small systematic discrepancy from other calibrations (\citealt{kewley2008metallicity}. It uses the line combinations of ([O~III]$\lambda$5007/H$\beta$)/([N~II]$\lambda$6584/H$\alpha$). 

{\it R$_{23}$}: The R$_{23}$-based metallicity diagnostics ($R_{23}$ = ([O~II]$\lambda$$\lambda$3727,3729 + [O~III]$\lambda$$\lambda$4959,5007)/H$\beta$) are one of the commonly used calibrations due to its insensitivity with ionisation parameter. We use the calibration investigated by \citet{pilyugin2005oxygen}. However, the independence only holds at \logOH $>$ 8.5 (\citealt{kewley2019understanding}).

\section{Description on the dust evolution model (\texttt{BEDE})}
\label{app:model_description}
This section specifies how the selected variables (dust-related parameters) contribute to the dust mass build-up/destruction over time.

The generic equation for the dust mass evolution in this model is as follows (Equation 10 in \citealt{devis2021bede}):
\begin{align}
    \begin{split}
        & \frac{dM_{\rm dust}}{dt} = \int_{m_t}^{m_U}([m - m_{R(m)}]~Z(t-\tau_{\rm m}) + mp_{\rm Z}) \\
        & ~~~~~~~~~~~~~~~~~~~\times C_{\rm dust}(m) \times \psi (t - \tau_{\rm m})\phi(m)dm \\
        & ~~~~~~~~~~~~~~~~~~~+ DGR_{I}I(t) - DGR(t)(O(t) + \psi(t)) + R_{M_{\rm dust}}(t) \\
        & ~~~~~~~~~~~~~~~~~~~+ f_{\rm c} f_{\rm dis} M_{\rm dust} \tau_{\rm gg,cloud}^{-1} + (1 - f_{\rm c}) M_{\rm dust} \tau_{\rm gg, dif}^{-1} \\
        & ~~~~~~~~~~~~~~~~~~~- (1 - f_{\rm c}) M_{\rm dust} \tau_{\rm dest}^{-1} \\
        & ~~~~~~~~~~~~~~~~~~~- (1 - f_{\rm c}) (1 - f_{\rm Si}) M_{\rm dust} \tau_{\rm frag}^{-1}.
    \end{split}
    \label{eq:dustevolv}
\end{align}
The integral form presents the \Mdust\ expelled by stars and SNe, which ranges from $M_{t}$, the lowest mass at a star's end of life at time $t$, to $M_{U}$, the upper mass limit of the star ($= 120$~\Msol). Specifically, 
$m_{R(m)}$ is the remnant mass of a star (\citealt{ferreras2000probing}), and $\tau_{\rm m}$ is the lifetime of stars with initial mass $m$, which is derived from \citet{schaller1992new}. $Z$ in this question means the metal mass fraction over gas mass. The $mp_{\rm Z}$ presents the metal yield from SN or AGB. The second term ($C_{\rm dust} (m))$ indicates the condensation efficiency of \Mdust, from low-to-intermediate mass stars ($m < 8$\Msol; set to 0.15 following the experiment in \citet{devis2021bede} where it is shown that its change has a negligible effect on the model) and SNe. For $m > 8$\Msol, the SN dust yield table (\citealt{todini2001dust}; \citealt{rowlands2014dust}) is used and the $SN_{\rm red}$ factors it as
\begin{equation}
    C_{\rm dust}(m > 8M_{\odot}) = 1/SN_{\rm red} \times C_{\rm dust, SN} [M_{\odot}/SN].
\end{equation}
$\psi$ is the star formation rate, and the $\tau_{\rm m}$ represents the lifetime of stars with initial mass $m$. The third term reflects the inflow, outflow and recycled dust components from the outflow. $DGR_{I}$ stands for the DGR of the inflows, and we set it to zero. $DGR_{O}$ is the DGR value of the outflows at the time $t$. We take only into account the low-velocity ($V_{\rm out}$ < 150~\kms) outflow components for the recycled \Mdust, $R_{M_{\rm dust}}$.

The fourth row of Equation~\ref{eq:dustevolv} brings the grain growth from dense clouds and diffuse clouds individually. The $f_{\rm c}$ is the mass fraction of cold dense clouds out of total ISM mas, and $f_{\rm dis}$ is the factor of dust from grain growths surviving from gas cloud dissociation. The grain growth timescale of each dense/diffuse cloud is described below.
\citet{devis2021bede}):
\begin{equation}
    \tau_{\rm gg,dif} = k_{\rm gg,dif} Z_{0} \left(1 - \frac{M_{\rm dust}}{M_{\rm metals} \times f_{\rm dif}} \right),
\end{equation}
\begin{equation}
    \tau_{\rm gg,cloud} = k_{\rm gg,cloud} Z_{0} \frac{SFR}{M_{gas}} \left(1 - \frac{M_{\rm dust}}{M_{\rm metals} \times f_{\rm cloud}} \right),
\end{equation}
where $k_{\rm gg,cloud}$ and $k_{\rm dif}$ controls the grain growth timescale in the dense and diffuse clouds, respectively. The $Z_{0}$ is the normalised metallicity by MW metallicity. In our model run, we fix the $k_{\rm gg,dif}$ to 5, and the $k_{\rm gg,cloud}$ is allowed to vary.

The dust destruction term by SN shocks can be described below:
\begin{equation}
    \tau_{\rm destr}^{-1} = 135 M_{\rm gas}^{-1} R_{\rm SN} M_{\rm destr},
\end{equation}
where $R_{\rm SN}$ is the SN rate, and we set the $M_{\rm destr}$ to 15~\Msol\ per SN in our model run.
\textit{$k_{\rm frag}$}, which regulates the timescale of the photofragmentation of a-C:H/a-C grains by UV radiation and given as:
\begin{equation}
    \tau_{\rm frag}^{-1} = k_{\rm frag} SSFR,
\end{equation}
where SSFR represents the diffuse UV radiation field.

\section{Multi-wavelength photometric data}
\renewcommand{\arraystretch}{0.8}
\begin{table}
    \centering
    \begin{tabular}{c|c|c|c|c}
        \hline
        \hline
        Instrument & Band & $\lambda_{\rm eff}$ & Pixel width (optical) \\
         & & & / Resolution FWHM\\
        \hline
        GALEX & FUV & 152.8~nm & 4.3\arcsec \\
        GALEX & NUV & 256.6~nm & 5.2\arcsec \\
        CTIO 0.9m/1.0m/1.5m & U & 365.0~nm & 0.4\arcsec\ /0.3\arcsec\/ /0.2\arcsec \\
        CTIO 0.9m/1.0m/1.5m & B & 445.0~nm & 0.4\arcsec\ /0.3\arcsec\/ /0.2\arcsec \\
        CTIO 0.9m/1.0m/1.5m & V & 464.9~nm & 0.4\arcsec\ /0.3\arcsec\/ /0.2\arcsec \\
        CTIO 0.9m/1.0m/1.5m & R & 551.0~nm & 0.4\arcsec\ /0.3\arcsec\/ /0.2\arcsec \\
        CTIO 0.9m/1.0m/1.5m & I & 658.0~nm & 0.4\arcsec\ /0.3\arcsec\/ /0.2\arcsec \\
        KPNO 0.9m & U & 364.5~nm & 0.43\arcsec \\
        KPNO 0.9m & B & 431.2~nm & 0.43\arcsec \\
        KPNO 0.9m & V & 543.3~nm & 0.43\arcsec \\
        KPNO 0.9m & R & 645.8~nm & 0.43\arcsec \\
        KPNO 0.9m & I & 833.3~nm & 0.43\arcsec \\
        SDSS & u & 353.1~nm & 1.3\arcsec \\
        SDSS & g & 462.7~nm & 1.3\arcsec \\
        SDSS & r & 614.0~nm & 1.3\arcsec \\
        SDSS & i & 746.7~nm & 1.3\arcsec \\
        SDSS & z & 888.7~nm & 1.3\arcsec \\
        2MASS & J & 1.24$\rm \mu$m & 2.0\arcsec \\
        2MASS & H & 1.66$\rm \mu$m & 2.0\arcsec \\
        2MASS & K$\rm _s$ & 2.16$\rm \mu$m & 2.0\arcsec \\
        WISE & W1 & 3.4$\rm \mu$m & 6.1\arcsec \\
        IRAC & I1 & 3.6$\rm \mu$m & 1.66\arcsec \\
        IRAC & I2 & 4.5$\rm \mu$m & 1.72\arcsec \\
        WISE & W2 & 4.6$\rm \mu$m & 6.4\arcsec \\
        IRAC & I3 & 5.8$\rm \mu$m & 1.88\arcsec \\
        IRAC & I4 & 8.0$\rm \mu$m & 1.98\arcsec \\
        WISE & W3 & 12$\rm \mu$m & 6.5\arcsec \\
        WISE & W4 & 22$\rm \mu$m & 6.5\arcsec \\
        MIPS & M1 & 24$\rm \mu$m & 6\arcsec \\
        PACS & Blue & 70$\rm \mu$m & 9\arcsec \\
        PACS & Green & 100$\rm \mu$m & 10\arcsec \\
        PACS & Red & 160$\rm \mu$m & 13\arcsec \\
        SPIRE & PSW & 250$\rm \mu$m & 18\arcsec \\
        SPIRE & PMW & 350$\rm \mu$m & 25\arcsec \\
        \hline
        \hline
    \end{tabular}
    \caption{Main characteristics of multi-wavelength data used for this study. GALEX: \citet{morrissey2007calibration}; CTIO: \url{https://noirlab.edu/science/programs/ctio}; KPNO: \url{https://noirlab.edu/science/index.php/programs/kpno/filters/wiyn-09}; SDSS: \citet{doi2010photometric}; 2MASS: \citet{cohen2003spectral}; WISE: \url{https://wise2.ipac.caltech.edu/docs/release/allsky/expsup}; IRAC: \url{https://irsa.ipac.caltech.edu/data/SPITZER/docs/irac/iracinstrumenthandbook/IRAC_Instrument_Handbook.pdf}; MIPS: \url{https://irsa.ipac.caltech.edu/data/SPITZER/docs/mips/mipsinstrumenthandbook/MIPS_Instrument_Handbook.pdf}; PACS: \url{https://www.cosmos.esa.int/documents/12133/996891/PACS+Explanatory+Supplement/} SPIRE: \citet{griffin2010herschel}}
    \label{tab:ancillary}
\end{table}

\begin{table*}
    \centering
    \begin{tabular}{c|c|c|c}
        \hline
        \hline
        Galaxy & Ancillary data$^{a}$ & Ref.$^{b}$ \\
        \hline
        NGC~625 & FN, UBVR, JHK, W1/2/3/4, I1/2/3/4, P1/2/3, S1/2 & D09, Co14, Cu13, J03, M13 \\
        NGC~1512 & FN, BVRI, JHK, W1/2/3/4, M1, P1/2/3, S1/2 & G07, Ke03, Co14, Cu13, J03, D09, Ke11 \\
        NGC~1566 & FN, BVRI, JHK, W1/2/3/4, I1/2/3/4, M1, P1/2/3, S1/2 & G07, Ke03, J03, Cu13, Ke03, Cl13 \\
        NGC~1705 & FN, BVRI, JHK, W1/2/3/4, I1/2/3/4, M1, P1/2/3, S1/2 & G07, Co14, J03, Cu13, D09, M13 \\
        NGC~3521 & FN, ugriz, JHK, W1/2/3/4, I1/2, P1/2/3, S1/2 & G07, Br14, J03, Cu13, D09, Ke11 \\
        NGC~4536 & FN, ugriz, JHK, W1/2/3/4, I1/2, P1/2/3, S1/2 & G07, Br14, H03, J03, Cu13, S10, Ke11 \\
        NGC~5236 & FN, UBVR, JHK, W1/2/3/4, I1/2/3/4, M1, P1/3, S1/2 & Bi10, Ku00, J03, Cu13, Be11 \\
        NGC~5253 & FN, UBVR, JHK, W1/2/3/4, I1/2/3/4, P1/2/3, S1/2 & G07, Co14, J03, D09, M13 \\
        NGC~6822 & FN, UBV, JHK, W1/2/3/4, I1/2/3/4, M1, P1/2/3, S1/2 & H10, Ke03, J03, Cu13, M13 \\
        NGC~7793 & FN, UBVR, JHK, W1/2/3/4, I1/2/3/4, M1, P1/2/3, S1/2 & G07, Co14, J03, Cu13, D09, Ke11 \\
        Sextans~A & FN, UBVR, JHK, I1/2, M1, P1/2/3, S1/2/3 & G07, Co14, D09, J03, Cl13 \\
        \hline
        \hline
    \end{tabular}
    \caption{Ancillary multi-wavelength data used in this study. \\
    $^{a}$ FN = $GALEX$ Far/Near-UV, UBVRI = $CTIO$/$KPNO$ U/B/V/R/I bands, ugriz = $SDSS$ u/g/r/i/z bands, JHK = $2MASS$ J/H/K bands, W1/2/3/4 = $WISE$ 3.4/4.6/12/22 $\mu$m, I1/2/3/4 = $Spitzer-IRAC$ 3.6/4.5/5.8/8.9 $\mu$m, M1 = $Spitzer-MIPS$ 24 $\mu$m, P1/2/3 = $Herschel-PACS$ 70/100/160 $\mu$m, S1/2 = $Herschel-SPIRE$ 250/350 $\mu$m. \\
    $^{b}$ Reference: D09 = \citet{dale2009spitzer}, Co14 = \citet{cook2014spitzer}, Cu13 = \citet{cutri2021vizier}, J03 = \citet{jarrett20032mass}, M13 = \citet{madden2013overview}, G07 = \citet{gildepaz2007galex}, K03 = \citet{kennicutt2003sings}, K11 = \citet{kennicutt2011kingfish}, Br14 = \citet{Brown2014atlas}, S10 = \citet{Sheth2010spitzer}, Bi10 = \citet{bigiel2010tightly}, Ku00 = \citet{kuchinski2000comparing}, Be11 = \citet{bendo2012investigations}, H10 = \citet{hunter2010galex}, Cl13 = \citet{clark2018dustpedia}}
    \label{tab:ancillary_ref}
\end{table*}

\section{Radial profile of ISM surface density, DGR, and \logOH\ and quantity maps}
\label{app:eachsample}

\begin{figure*}
    \centering

    \includegraphics[width=0.73\textwidth]{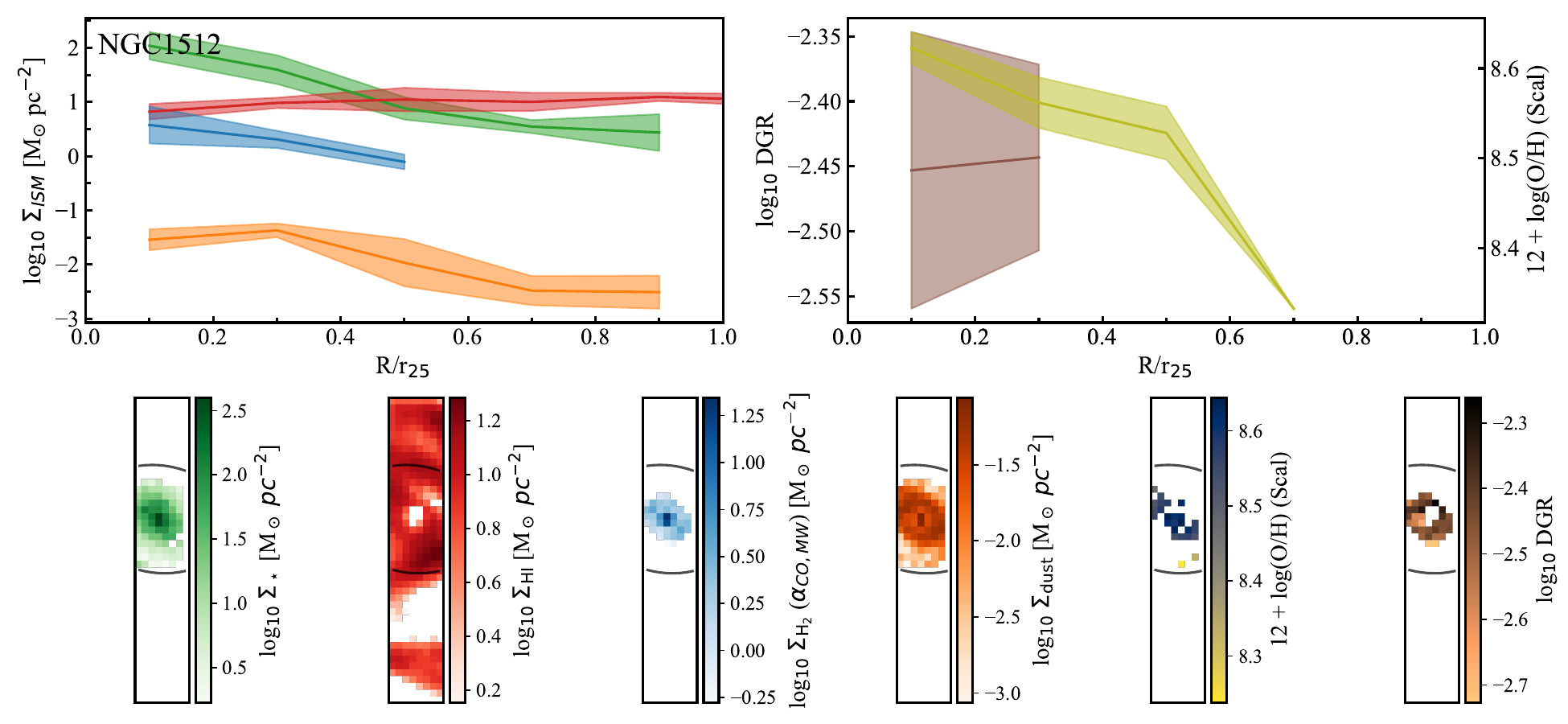}

    \caption{Radial profile of ISM surface density ($\Sigma_{\rm ISM}$), DGR, and Z$_{gas}$ of NGC~1512 (upper panels). Upper left panel: Radial profiles of $\Sigma_{\rm \star}$ (green), $\Sigma_{\rm H_2}$ (blue), $\Sigma_{\rm HI}$ (red), $\Sigma_{\rm dust}$ (orange) in log scale Upper right panel: Radial profiles of \logOH\ (Scal; olive) and DGR (brown) in log scale. For the radius bin with no more than one pixel, we show the profile with dots and connect them using dashed lines. Lower panels: The quantity maps of each parameter with the r$_{25}$ presented with a black solid line. The radial profiles and the quantity maps for other galaxies are available in the online version of the article.}
    \label{fig:NGC1512}
\end{figure*}

We construct a radius map of each galaxy using the following equation:
\begin{align}
    \begin{split}
    & R (x,y) = \sqrt{{\mathrm cos}^2\theta + {\mathrm sin}^2\theta}\,,\\
    & {\rm where}\\
    & \mathrm{cos}~\theta = - (x - x_{c}) \times {\mathrm sin}~(PA) + (y - y_{c}) \times {\mathrm cos}~(PA),\,\\
    & {\rm and}\\
    & \mathrm{sin}~\theta = (- (x - x_{c}) \times {\mathrm cos}~(PA) + (y - y_{c}) \times {\mathrm sin}~(PA) / {\mathrm cos}~(i),
    \label{eq:radius}
    \end{split}
\end{align}
where $R$ represents the radius, $x$ and $y$ are the pixel coordinates, $x_c$ and $y_c$ are the centre of the galaxy in pixel locations, $PA$ is the position angle, and $i$ is the inclination of the galaxy, as listed in Table~\ref{tab:sample}. The inclination effect has been corrected for the ISM surface density maps and radial profiles of each galaxy using the information in Table~\ref{tab:sample}. The radius is normalized by $r_{25}$ and binned into several bins with a step size of 0.2 $r_{25}$ and the radial profile of each parameter is shown by choosing the median value of physical parameters within the defined radius range with dashed lines. The error bars of the radial profile are from the standard deviation of the quantity at each radius bin. We note that the H{\sc i} mass and H$_2$ gas mass have been corrected for the helium abundance. The H$_2$ gas surface density is derived using the constant \alphacomw\ as is DGR. Figure~\ref{fig:NGC1512} provides the radial profiles of each ISM component as an example. Other galaxies are available in the online version of this article.

\bsp	
\label{lastpage}
\end{CJK}
\end{document}